\documentclass[]{raa}
\usepackage{graphicx,times}
\usepackage{subfigure}
\usepackage{amssymb}
\usepackage{threeparttable}
\usepackage{epstopdf}
\usepackage{hyperref}
\usepackage{caption}
\usepackage{booktabs}
\usepackage{multirow}
\usepackage{array}
 

\begin{document}

    \title{Prompt Optical Emission from Gamma-ray Bursts with Non-single Timescale Variability of Central Engine Activities }

    \volnopage{Vol.0 (200x) No.0, 000--000}
    \setcounter{page}{1}

    \author{Si-Yao Xu
       \inst{1}
    \and Zhuo Li
        \inst{1,2}}
    \institute{Department of Astronomy and Kavli Institute for Astronomy and Astrophysics, Peking University, Beijing 100871, China; \\
         \and
              Key Laboratory for the Structure and Evolution of Celestial
       Objects, Chinese Academy of Sciences, Kunming 650011, China.  \\ {\it Email: syxu@pku.edu.cn}}

\abstract{The complete high-resolution lightcurves of Swift GRB
080319B present an opportunity for detailed temporal analysis of the
prompt optical emission. With a two-component distribution of
initial Lorentz factors, we simulate the dynamical process of the
ejected shells from the central engine in the framework of the
internal shock model. The emitted radiation are decomposed into
different frequency ranges for a temporal correlation analysis
between the lightcurves in different energy bands. The resulting
prompt optical and gamma-ray emission show similar temporal
profiles, both as a superposition of a slow variability component
and a fast variability component, except that the gamma-ray
lightcurve is much more variable than its optical counterpart. The
variability features in the simulated lightcurves and the strong
correlation with a time lag between the optical and gamma-ray
emission are in good agreement with the observations of GRB 080319B.
Our simulations suggest that the variations seen in the lightcurves
stem from the temporal structure of the shells injected from the
central engine of gamma-ray bursts. The future high temporal
resolution observations of prompt optical emission from GRBs, e.g.,
by UFFO-Pathfinder and SVOM-GWAC, provide a useful tool to
investigate the central engine activity. }
      \authorrunning{S.Y. Xu \& Z. Li}
 \titlerunning{Prompt optical emission from GRBs}
 \maketitle

\section{Introduction}
\label{sect:intro}

Gamma-ray bursts (GRBs) are believed to be produced by the
relativistic jets released from the compact central engines, however
the composition of jets and the energy dissipation and radiation
mechanism at work are still far from clear. In the widely used
internal shock model (\cite{Rees94}), the energy dissipation in GRBs
is caused by collisions between different parts of the unsteady
outflow. These collisions produce shocks which accelerate electrons
and generate magnetic field, and the GRB prompt emission is produced
by the synchrotron radiation from the accelerated electrons. The
internal shock model can generally match the gamma-ray properties of
GRBs. For typical model parameters, the internal shock synchrotron
model can naturally explain the complexity of GRB light curves
(\cite{Koba97}), the spectral break energy around MeV range, and the
high energy photon index of $\beta\sim-2$ (see review of
\cite{waxman03})\footnote{Note, there are also the other energy
dissipation model involving magnetic energy dissipation by
reconnection and turbulence etc. (e.g., \cite{usov94,
thompson94,Lyu03,narayan,zhangyan}) and the other radiation
mechanism, e.g., thermal radiation (\cite{mr00,beloborodov10}).}.
Moreover the "fast cooling problem" of the low energy photon index
can also be reconciled by involving postshock magnetic field decay
in the internal shock model (\cite{peer06,zhaoli13}).

The observations of GRB prompt emission outside the MeV energy range
will further help to diagnose the jet properties and the central
engine activity. The Fermi-LAT observations of bright GRBs reveal
that the GRB emission in GeV range also shows short timescale,
$\la1$~s, variabilities in both long and short GRBs (e.g.,
\cite{Ad09a,Ad09b}), implying the similar origin related to MeV
emission. However, the temporal delay of GeV emission relative to
MeV one implies larger radii of GeV emission than that of MeV ones
(\cite{li10}). Moreover, the fact that the GeV emission is dominated
by MeV one supports that the radiation mechanism at work for MeV
emission is synchrotron radiation other than inverse-Compton
scattering (\cite{wang09}; see also \cite{derishev02,piran09}). On
the other hand, the prompt optical emission during the gamma-ray
emission is detected in some GRBs (e.g., GRB 990123, \cite{Aker99};
GRB 041219A, \cite{Vest05}; GRB 051109A and GRB 051111,
\cite{Yost07}; GRB 061121, \cite{Page07}; see also
\cite{kobayashi13}, and references therein). The bright optical
emission also implies that the radius of the optical emission region
is larger than MeV one, in order to avoid the synchrotron self
absorption (Li \& Waxman~\cite{lw08,fan09,shenzhang,zou}). However
the temporal optical properties in short time scale, $\la1$~s, are
not clear due to the low time resolution in optical observations.

By luck, the detection of the naked-eye burst, GRB 080319B, presents
the only and best-sampled case hitherto for analyzing the overall
temporal structure of the lightcurves of GRBs (\cite{Rac08};
Beskin et al.~\cite{Besk10}). The high (sub-second) temporal resolution data,
acquired from the onset of the optical transient to the end, makes
the burst possible for revealing the detailed structure of the
optical emission, shedding light on the behavior of the burst
internal engine (Beskin et al.~\cite{Besk10}). There are two main temporal
properties in the optical light curve of GRB 080319B. First, the
onset of the optical emission is delayed relative to the gamma-ray
one by $\sim10$s; Second, in the plateau phase the optical light
curve is correlated to gamma-ray one but with a time lag of
$\sim2$s. The former property may be due to the effect of
environment, e.g., the dust effect in the molecular cloud
(\cite{cui13}), while the latter is more likely tracking the time
history of the central engine, which is the focus of this paper.

As pointed out in Li \& Waxman~(\cite{lw08}), within the context of the internal
shock models, after the first generation collisions producing
gamma-ray emission, collisions continue to happen in larger and
larger radii with smaller and smaller relative velocities. These
"residual collisions" produce longer and longer wavelength
emission, which can avoid the strong synchrotron absorption in
gamma-ray emission region and produce strong optical emission as
observed. Because of the larger emission radii, the optical emission
is expected to systemically delay relative to the gamma-ray one.
Moreover the correlation between optical and gamma-ray light curves
in GRB 080319B implies a large timescale modulation in the central
engine activity.

In this paper we carry out numerical simulations of internal shocks
to produce multi-band, including gamma-ray and optical, light
curves, with focus on the effect of non-single timescale activities
of the central engines. In Section 2, we provide a general
description of the model we use. In Section 3, based on the model,
we perform several simulation tests with different initial Lorentz
factor distributions, following which we derive both optical and
gamma-ray lightcurves with different temporal structures. In Section
4, we compare the model results with the observational data of GRB
080319B. We discuss the implications on the central engine
activity in Section 5, and draw our conclusions in Section 6.

\section{Model}\label{sect:mod}
The GRB unsteady outflow can be approximated by a set of individual
shells with the initial position $r_i$, mass $m_i$, Lorentz factor
$\gamma_i$ and width $\Delta_i$ ($i=1,...,N$) at time $t=0$. The
initial widths can be taken to be the size of the source,
$\Delta_{1,...,N}=\Delta_0\approx10^6$cm. Given the initial
condition of the shells, the dynamical evolution afterward will be
totally fixed. Two neighboring shells $i$ and $i+1$ which satisfy
$v_{i+1}>v_i$ may collide at a time of $t+\Delta t$, where $\Delta
t=(r_i-r_{i+1})/(v_{i+1}-v_i)$, and the colliding radius is
$r_c=r_i+v_i\Delta t=r_{i+1}+v_{i+1}\Delta t$. The pair of shells
with smaller $\Delta t$ will collide first and merge into a new
shell.

For the pair of colliding shells, the velocity of center of momentum
(c.m.) is $\beta_{cm}=\sum\gamma_i\beta_im_i/\sum\gamma_im_i=\sum
m_i\sqrt{\gamma_i^2-1}/\sum\gamma_im_i$ and the relevant Lorentz
factor (LF) is $\gamma_{cm}=1/\sqrt{1-\beta_{cm}^2}$. They are also
the velocity and LF of the merged shell, respectively. In the case
of $\gamma_{i,i+1}\gg1$, the c.m. LF can be approximated by
\begin{equation}
  \gamma_{cm}=\sqrt{\frac{\gamma_im_i+\gamma_{i+1}m_{i+1}}{m_i/\gamma_i+m_{i+1}/\gamma_{i+1}}}.
\end{equation}

The shell may expand and change the shell width. Before collision at
$r_c$, the width of a shell is
$\Delta_i=\max(\Delta_i,~r_c/\gamma_{cm}^2)$ (\cite{guetta01}).
Because of compression by the shocks, the width of the merged shell
right after shock crossing, $\Delta_c$, is smaller than the sum of
the shell widths right before the collision
$\Delta_c<\Delta_i+\Delta_{i+1}$ (We derive $\Delta_c$ in the
appendix). After merging, the parameters for merged shell are the
position $r_c$, mass $m_i+m_{i+1}$, LF $\gamma_{cm}$ and width
$\Delta_c$. Collisions continue to occur, and they will only stop
when the shell velocities are increasing with radius after
enough collisions and momentum exchanges between shells.

The internal energy generated in this collision is
\begin{equation}
  E_{in}/c^2=(\gamma_im_i+\gamma_{i+1}m_{i+1})-\gamma_{cm}(m_i+m_{i+1})
\end{equation}
which is released by synchrotron radiation. The emission appears as
a pulse in the light curve. For simplicity we assume the shape of
the pulse as a rectangle with a width of $\delta
t=r_c/2\gamma_{cm}^2c$. So the luminosity of the light curve pulse
is $L=E_{in}/\delta t$. Actually, a pulse with fast rise and
slower decline in lightcurves can be produced due to the effect of
equal arrival time surface (e.g., \cite{Koba97,Shen05,Hua07}).
However, the time resolution of the observed lightcurves in the
prompt phase is not sufficient to tell the detailed pulse shapes.
Thus we adopt a simple shape for each pulse and this assumption does
not affect our temporal analysis of the lightcurves. In
observation, the starting time of the pulse is at a observer time of
\begin{equation}
                t_{\rm obs} = t-r_c/c .
\end{equation}
For the numerous collisions taking place between multiple shells, we
superimpose pulses produced by each collision according to their
time sequence.

Let us calculate the characteristic energy of the synchrotron
photons. The shell $i$'s velocity and LF in c.m. frame are
\begin{equation}
  \beta'=\frac{\beta_i-\beta_{cm}}{1-\beta_i\beta_{cm}},
  ~~\gamma'=\gamma_i\gamma_{cm}(1-\beta_i\beta_{cm}).
\end{equation}
The characteristic postshock electron LF is
\begin{equation}
  \gamma_e\approx(\gamma'-1)\epsilon_em_p/m_e+1,
\end{equation}
where $\epsilon_e$ is the fraction of postshock energy carried by
electrons. The magnetic field in the shock frame  is given by
\begin{equation}
  B=\sqrt{\frac{2\epsilon_BE_{in}}{\gamma_{cm}^2r_c^2\Delta_c}},
\end{equation}
where $\epsilon_B$ is the fraction of postshock energy carried by
magnetic field. Note here we use the shock-compressed width
$\Delta_c$ (derived in the appendix) to calculate $B$. The
synchrotron photon energy is then
\begin{equation}
  \epsilon_{syn}\approx\gamma_{cm}\gamma_{e}^2\frac{heB}{2\pi m_ec}=\gamma_{e}^2\frac{he}{2\pi m_ec}\sqrt{\frac{2\epsilon_BE_{in}}{r_c^2\Delta_c}}
\end{equation}
independent of $\gamma_{cm}$. The equipartition values
$\epsilon_e=\epsilon_B=1/3$ will be used in the following numerical
simulation.

For long wavelength emission, the synchrotron absorption may be
important. Following Li \& Waxman~(\cite{lw08}) to calculate the synchrotron
absorption frequency when $\epsilon_a>\epsilon_{syn}$, we have
\begin{equation}
  \epsilon_{a}=2(\epsilon_BE_{in})^{1/14}\gamma_{cm}^{4/7}\gamma'^{2/7}(m_i+m_{i+1})^{2/7}(cr_c)^{-5/7}\Delta_c^{-5/14}\rm
  keV,
\end{equation}
where a flat electron distribution of $p=2$ is used. The energy
spectrum will peak at $\epsilon=\max(\epsilon_{syn},\epsilon_a)$.
For simplicity let us assume a $\delta$ function for the spectrum,
i.e., the radiation is emitted at $\epsilon$.

In order to inspect the temporal correlation between  the prompt
optical and gamma-ray emission, the radiation produced by two shell
collisions is decomposed into different frequency ranges, gamma-ray
($>100$~keV), X-ray ($1-10$~keV) and optical emission ($1-10$~eV),
according to the energy band at which the synchrotron emission
peaks.

\section{Simulation tests}
\label{sect:Lit}

In order to show the effect of multi-timescale variabilities on the
dynamics of the outflow, we carry out simulations for both cases of
single- and two-timescale variabilities for comparison. We consider
a series of individual material shells $i = 1, 2, ...,N$, with total
shell number $N$, released in a duration of $T = 20$~s, so that the
interval between two nearby shells is $\tau_1=T/N$. The shells have
equal masses but different LFs. We carry four simulation tests as
below. For Test 1, we use $N=2000$, thus $\tau_1=10$~ms, and the
bulk LF of each shell follows
\begin{equation}
  \log\gamma_i=\log50+\xi_i\log10+(2\xi_i-1)\log2,
\end{equation}
where $\xi_i$ is the random number between zero and unity. This is
the case with only a single timescale of $\tau_1=10$~ms. For Test 2,
there are two timescales in the LF evolution. Beside the
$\tau_1=10$~ms variability ($N=2000$), there is an additional slow
modulation of $\tau_2=5$~s,
\begin{equation}
  \log\gamma_i=\log50+\xi_i\log10+\sin[\frac{2\pi}{\tau_2}(i-1)\tau_1-1.5]\log2.
\end{equation}
We also test Test 3 and 4 similar to Test 2 but with different
values of $\tau_1$ or $\tau_2$: $\tau_1=10$~ms and $\tau_2=3.3$~s in
Test 3; while $\tau_1=40$~ms (i.e., $N=500$) and $\tau_2=5$~s in
Test 4.

\begin{figure}
\begin{minipage}[t]{0.5\textwidth}
\centering
\includegraphics[width=6.5cm]{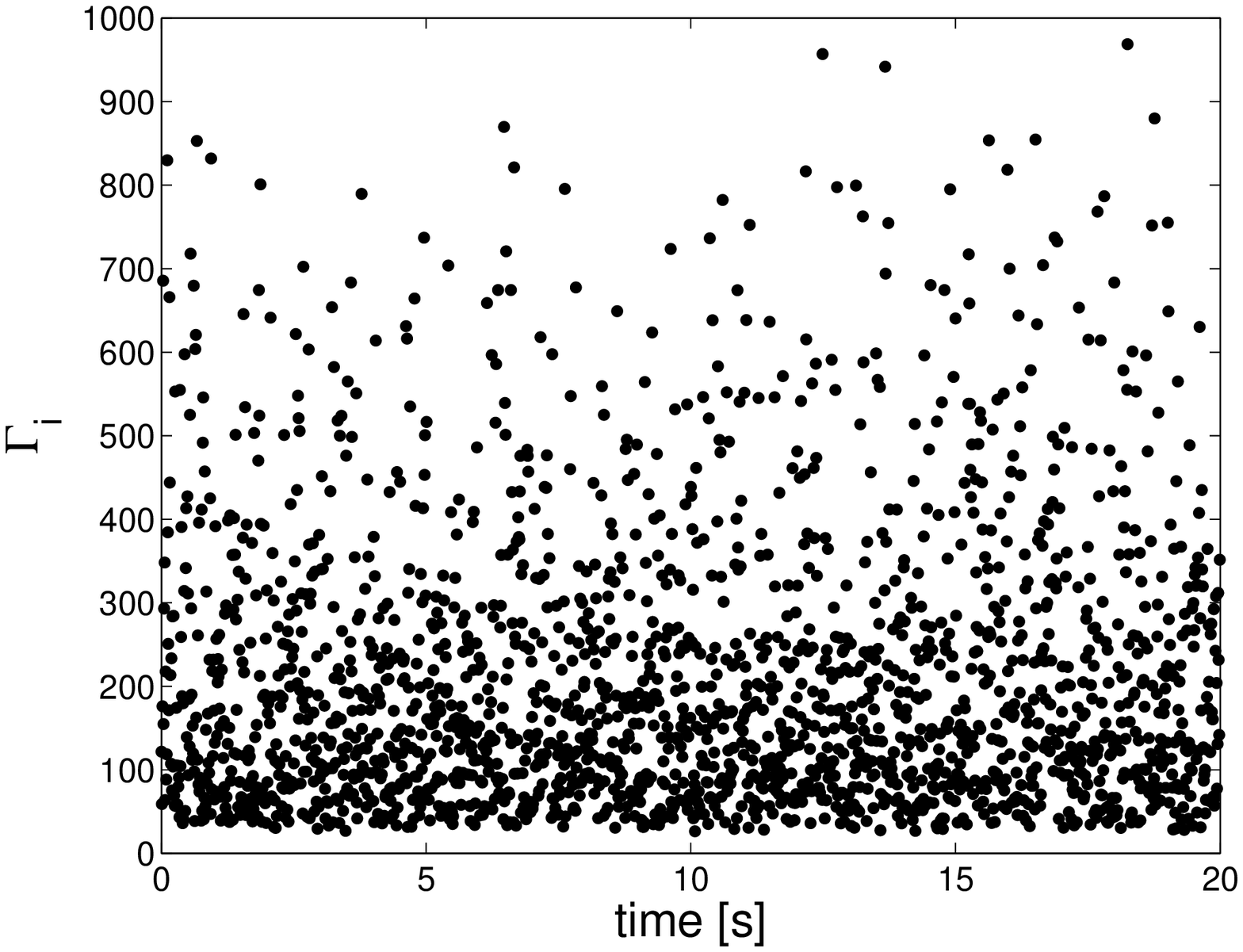}
\includegraphics[width=6.3cm]{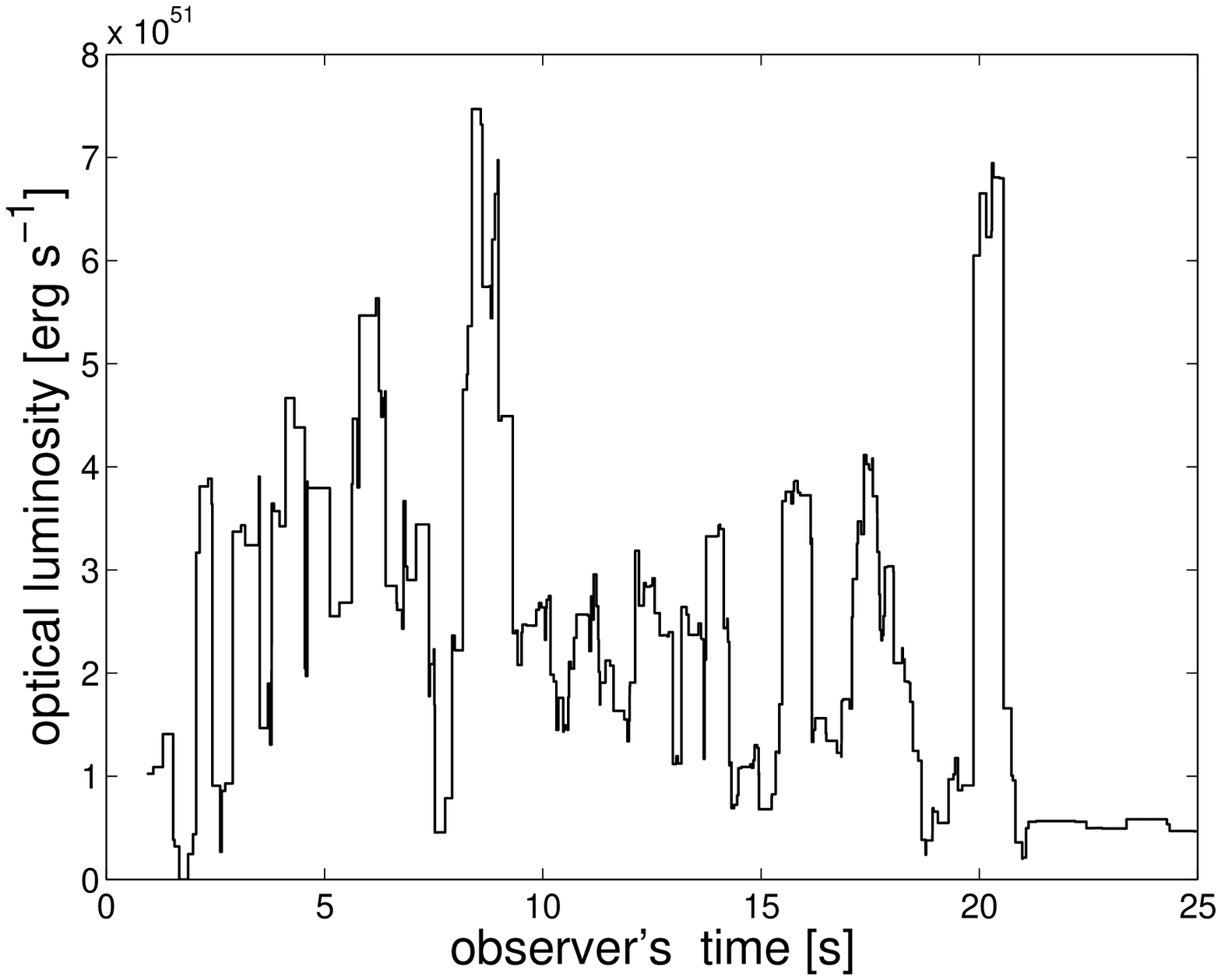}
\includegraphics[width=6.3cm]{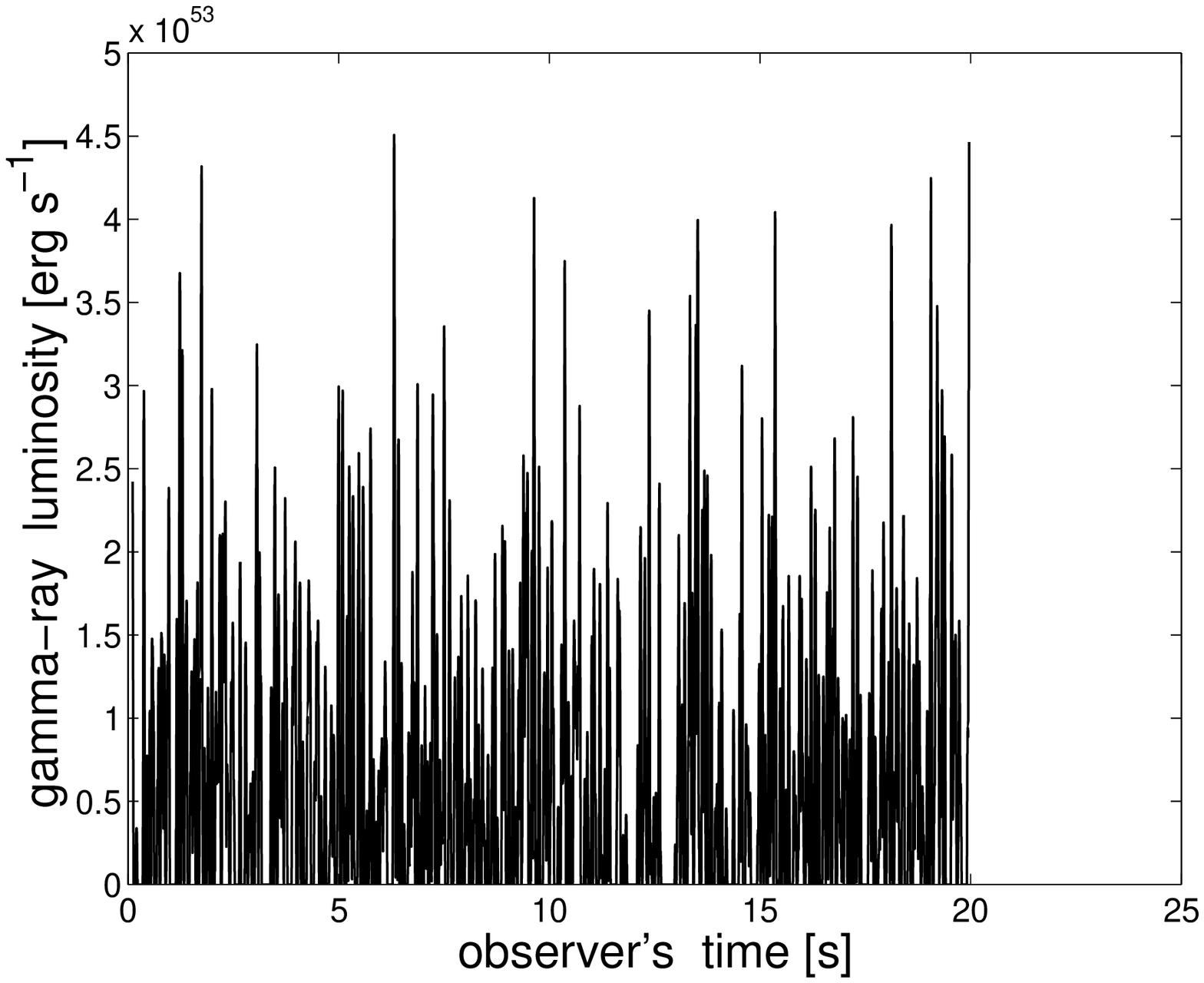}
\includegraphics[width=6.3cm]{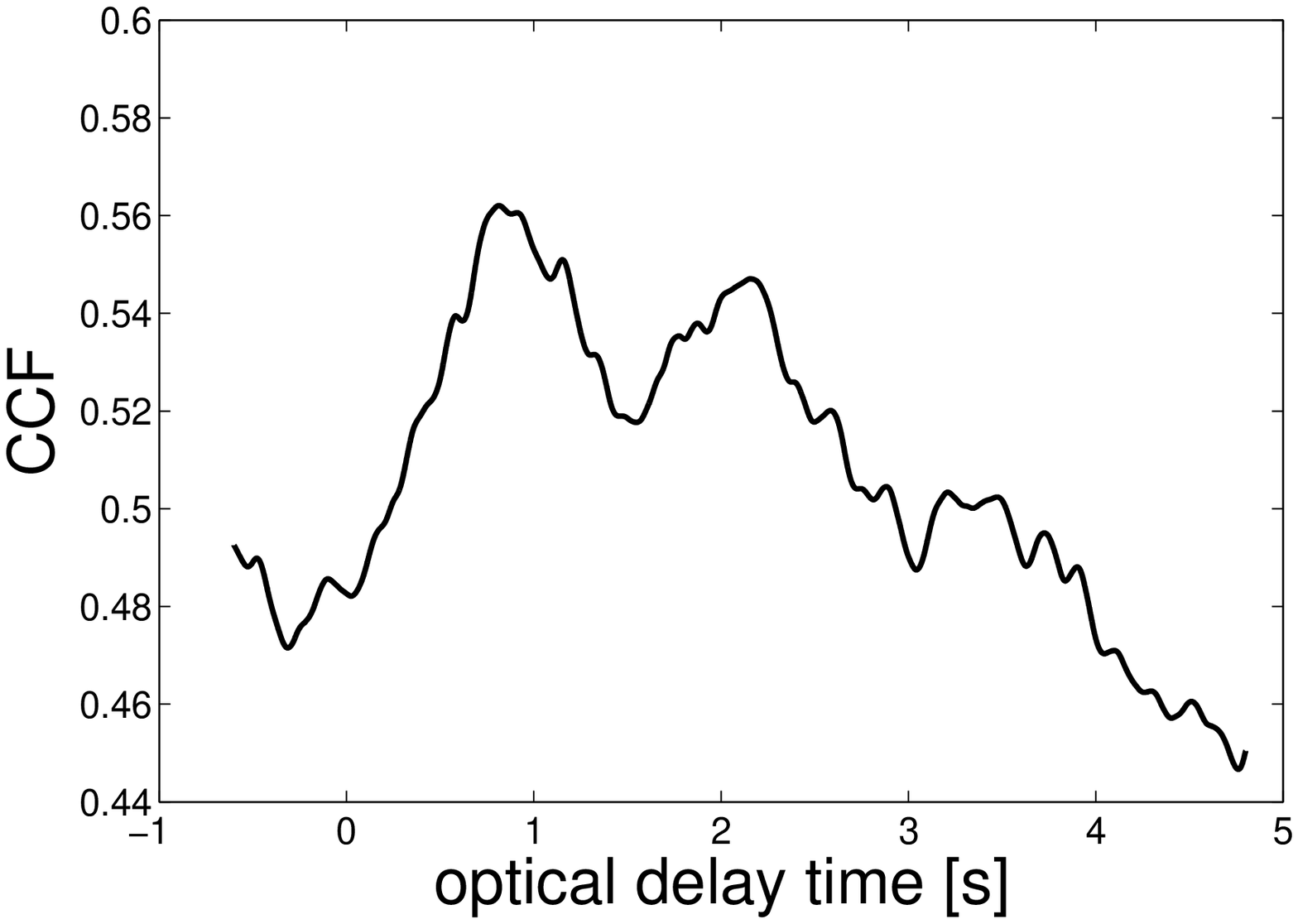}
\renewcommand{\captionlabelfont}{\bf}
   \captionsetup{labelsep=space}
   \caption{\small Results of Test 1. Top panel: the initial Lorentz factor distribution.
   Second panel: the optical light curve Third panel: the gamma-ray
lightcurve in 50 ms time bin. Bottom panel: the cross-correlation
function between the gamma-ray and optical lightcurves.
}\label{fig:test1}
\end{minipage}
\begin{minipage}[t]{0.5\textwidth}
\centering
\includegraphics[width=6.5cm]{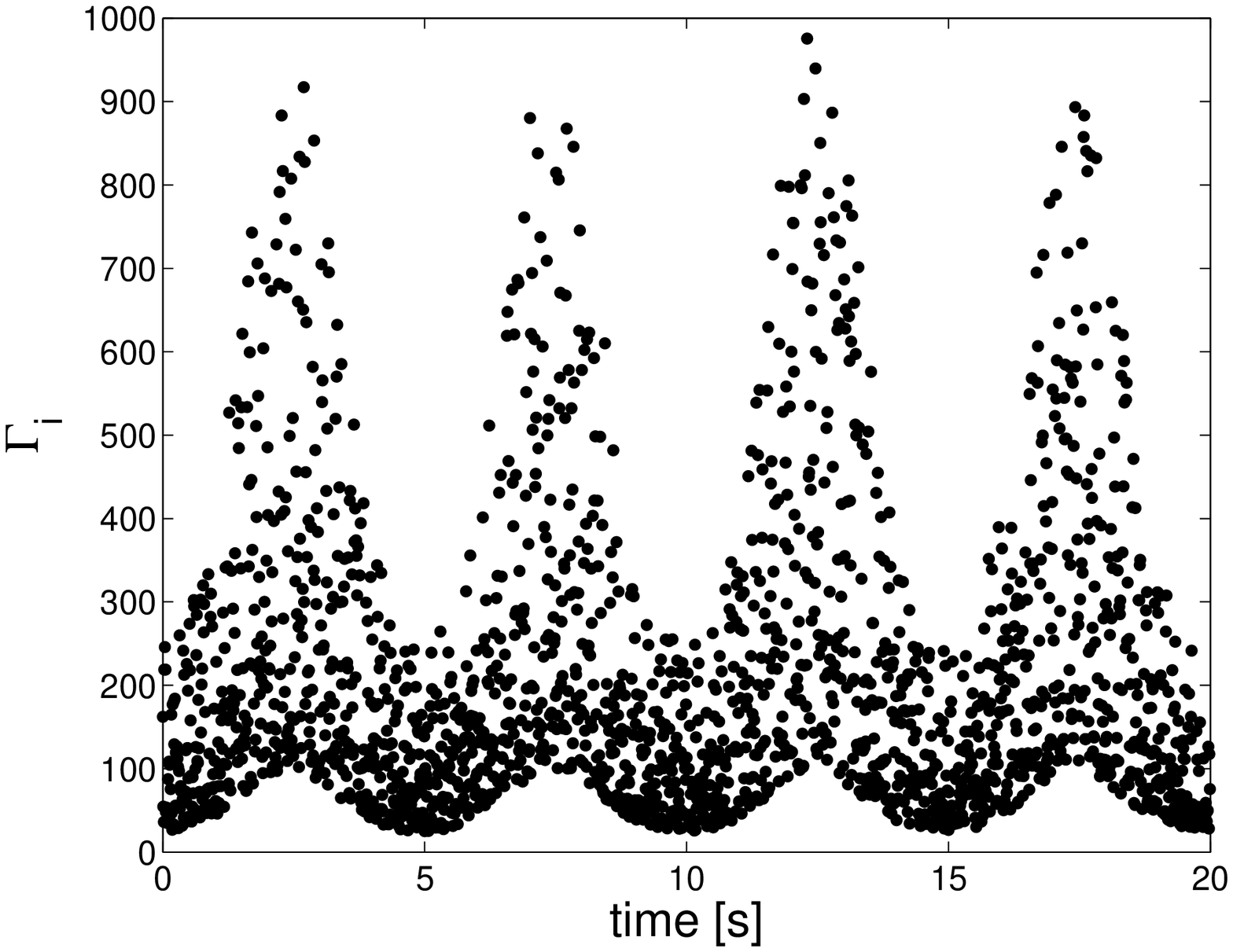}
\includegraphics[width=6.3cm]{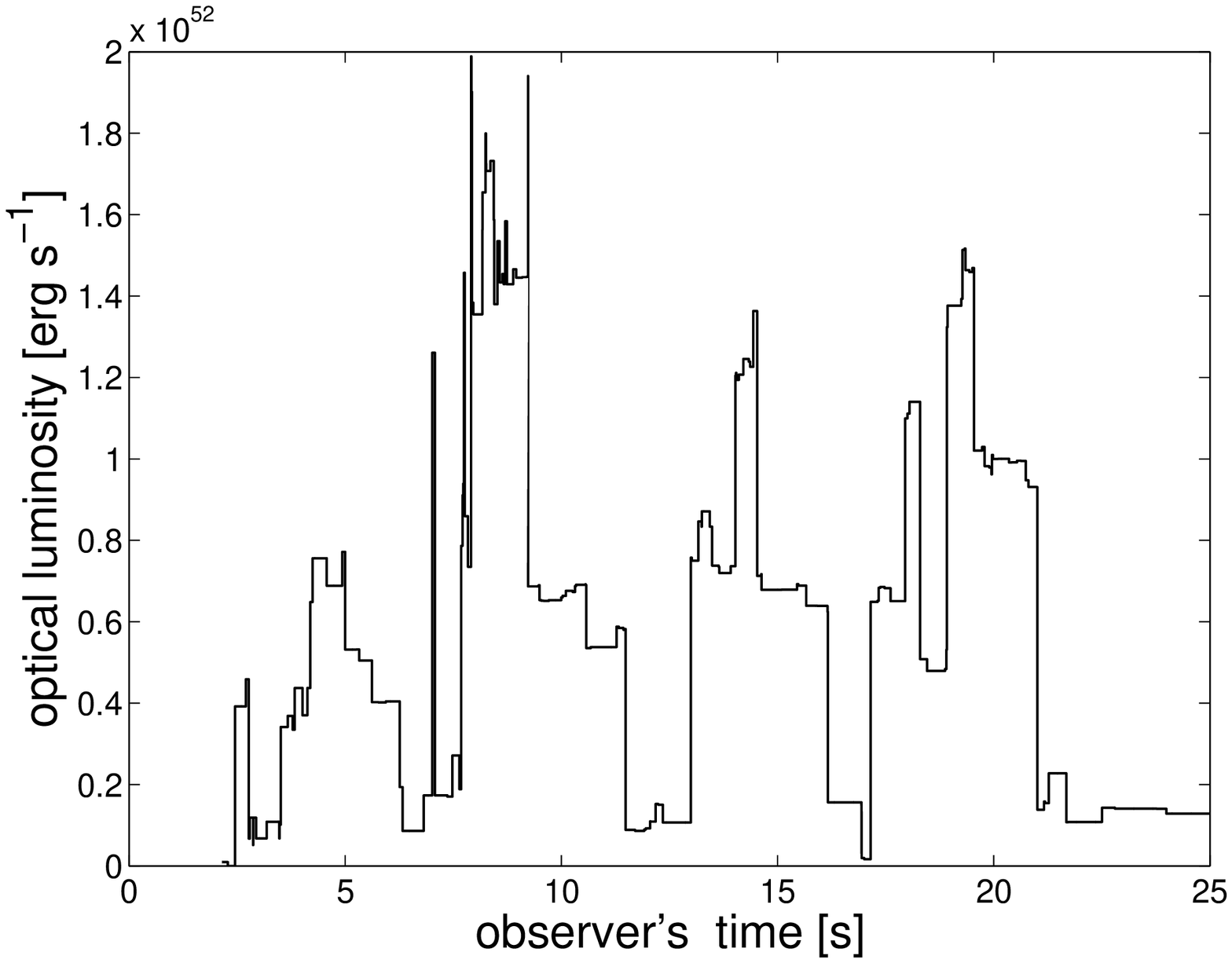}
\includegraphics[width=6.3cm]{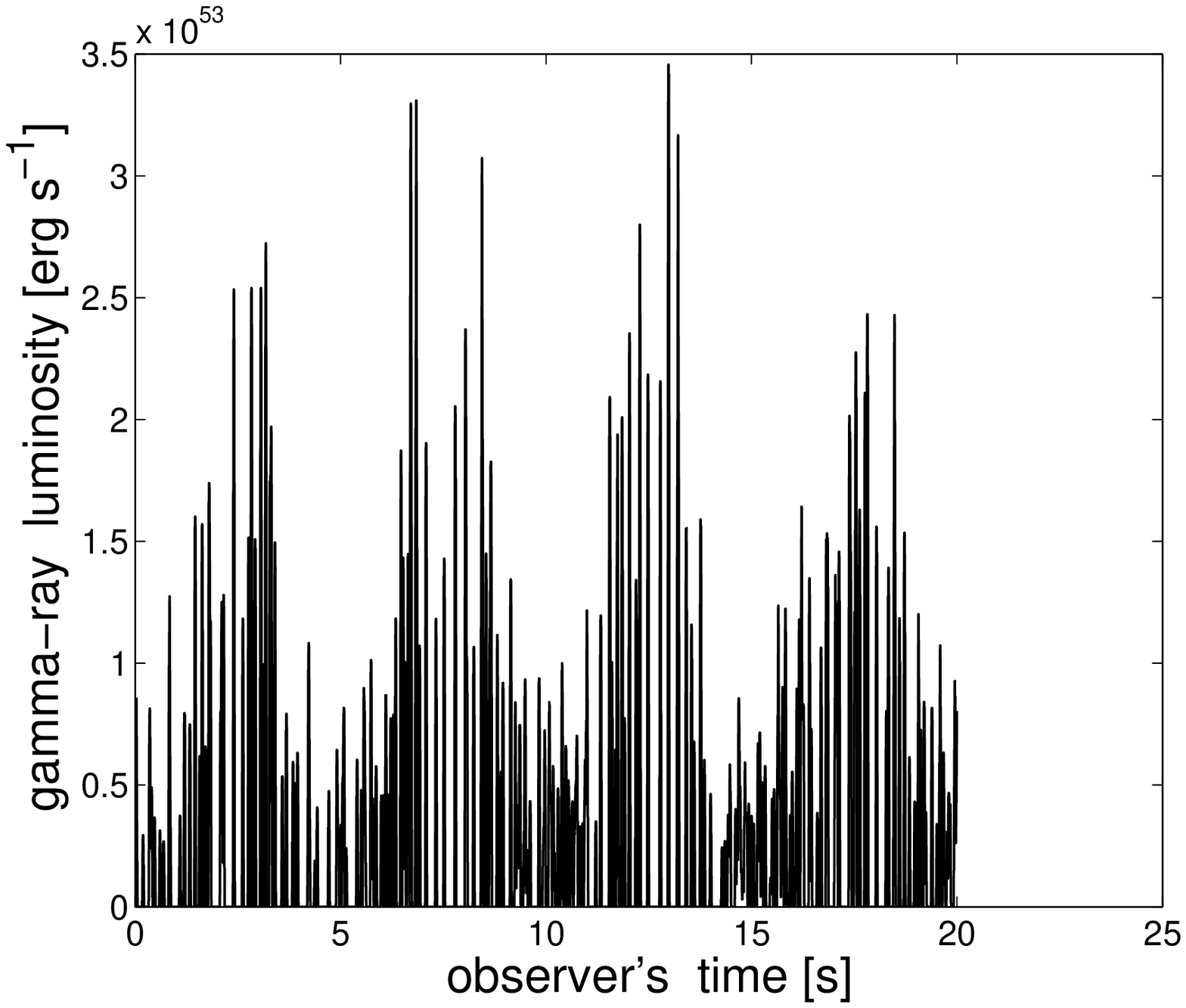}
\includegraphics[width=6.3cm]{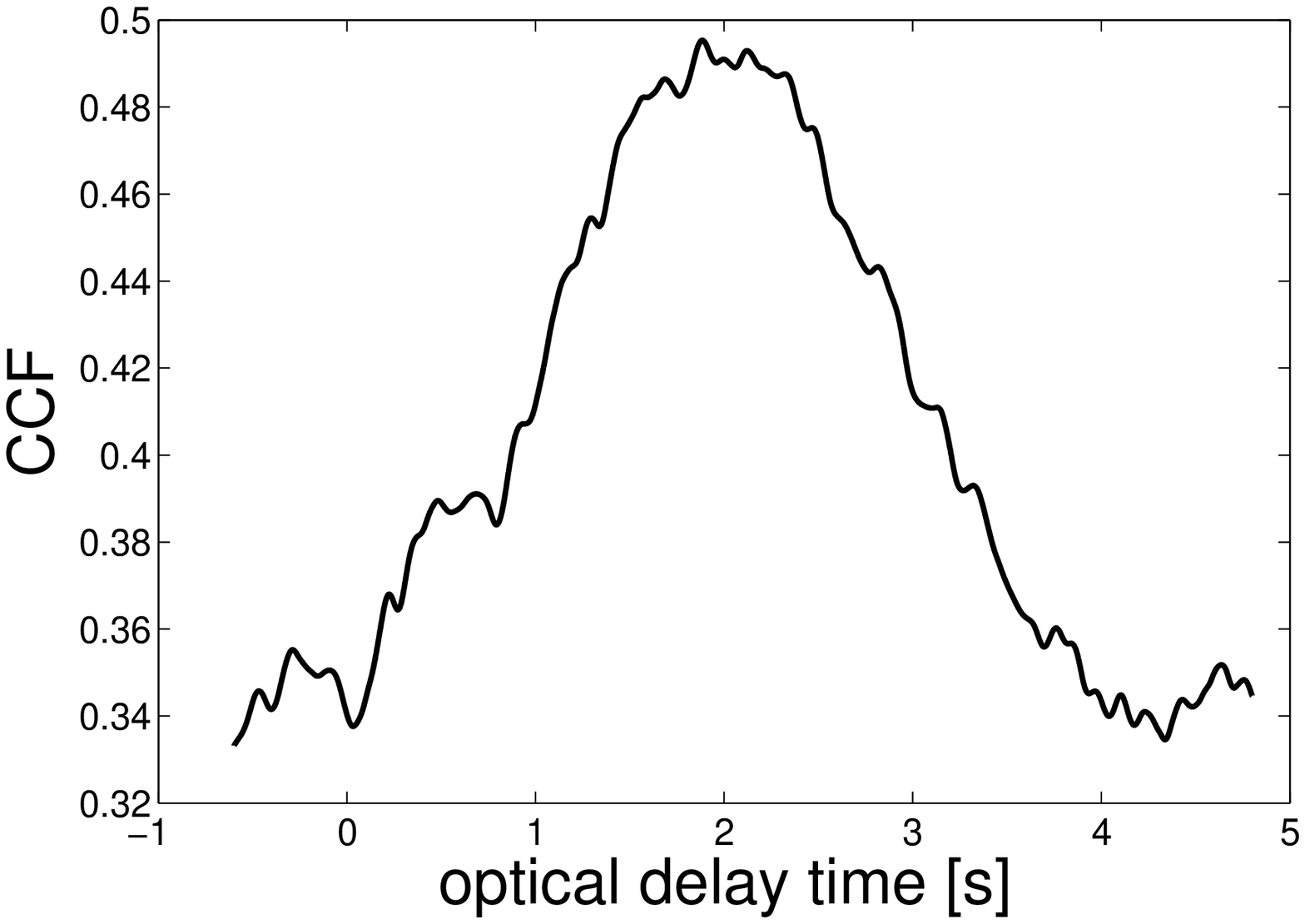}
\renewcommand{\captionlabelfont}{\bf}
\captionsetup{labelsep=space} \caption{\small Results of Test 2.
}\label{fig:test2}
\end{minipage}%
\end{figure}

\paragraph{Test 1.}
To examine the dependence of the temporal structure of lightcurves
on the  initial velocity variations inside the ejecta, we first
perform a simulation test using a succession of initial Lorentz
factors with only  a rapid random variability component. We consider
the case with uniformly distributed LFs presented in
Fig.~\ref{fig:test1}. Note that here we only display the first 20\,s
of the optical lightcurve for a clear comparison with the gamma-ray one.
It is straightforward to show that the
short-scale variability features of the LFs are imprinted on both
the optical and gamma-ray lightcurves (Fig.~\ref{fig:test1}). In
contrast to the broad periodic component seen in the following
simulation tests, there are only stochastic spikes existing in the
lightcurves.

In addition, Fig.~\ref{Fig:5tgr} shows the complete optical
lightcurves from our five simulation tests. Logarithmic luminosity
is used here due to the weak optical emission after $t_{\rm obs}
>20$~s. The optical emission clearly has a more variable temporal
profile in the early 20\,s than the rest part (also see Table 1),
i.e., the variability time in the late time is larger than the
early, $<20$s, emission.

\paragraph{Test 2.}
We start with a simple two-component Lorentz factor distribution of
the shells. Fig.~\ref{fig:test2} presents the distribution of the
initial LFs of the shells with indices from one to $N$. In our
calculations, the $N$th shell is the first shell emitted by the
inner engine, indicating the outer edge of the ejecta. The overall
duration of the burst is $T=20$\,s in the observer frame. The slow
and periodic variability existing in the LFs is on a time scale of
$\tau_2=5$\,s, and the overlapping rapid and irregular variations
have $\tau_1=10$\,ms time scale. Fig.~\ref{fig:test2} shows the
corresponding optical and gamma-ray lightcurves produced by a
Monte-Carlo simulation of the dynamic process of the colliding
shells we described in Section~\ref{sect:mod}.

By comparing these two lightcurves, we find they  both show a
superposition of two variability components: a slow periodic
component with a duration of $\sim$ 5\,s and a fast component with
stochastic short pulse widths. But in contrast to the smoother
profile in optical band, the gamma-ray lightcurve is obviously
highly variable with more rapid short-scale variabilities (see Table
1). To have a better comparison with the observed data, we rebinned
the optical and gamma-ray lightcurves with a 0.13 s and 50 ms bin
size, respectively (Beskin et al.~\cite{Besk10}), and performed a
cross-correlation analysis between them. Fig.~\ref{fig:test2}
displays the normalized cross-correlation sequence between the
optical and gamma-ray lightcurves as a function of optical delay
time.The correlation coefficient reaches its highest value ($\sim$ 0.50)
when the optical flux is delayed by 1.9 s with respect to the
gamma-ray emission. Since our results are just based on the simple
modeling of the internal shock dynamics, the time delay naturally
results from the evolution of Lorentz factor fluctuations of the
fast moving shells. At first, gamma-ray emission originates from the
outflow with highly variable LFs at a small radius. As the flow
moves forward to a larger radius, the variance of the Lorentz
factors of the remained shells decreases, leading to the decrease of
the radiated energy after collisions, and the characteristic
frequency of the emission as well (Li \& Waxman~\cite{lw08}).

The average ratio of the optical and gamma-ray fluxes indicates relatively bright optical
emission accompanying gamma-ray emission, which is consistent with
optical detections. The bright optical emission can be naturally
explained in the framework of the residual collisions model, since
optical emission can be produced at large radii, where the optical
depth to optical photons is low (Li \& Waxman~\cite{lw08}).

\begin{figure}
\begin{minipage}[t]{0.5\textwidth}
\centering
\includegraphics[width=6.5cm]{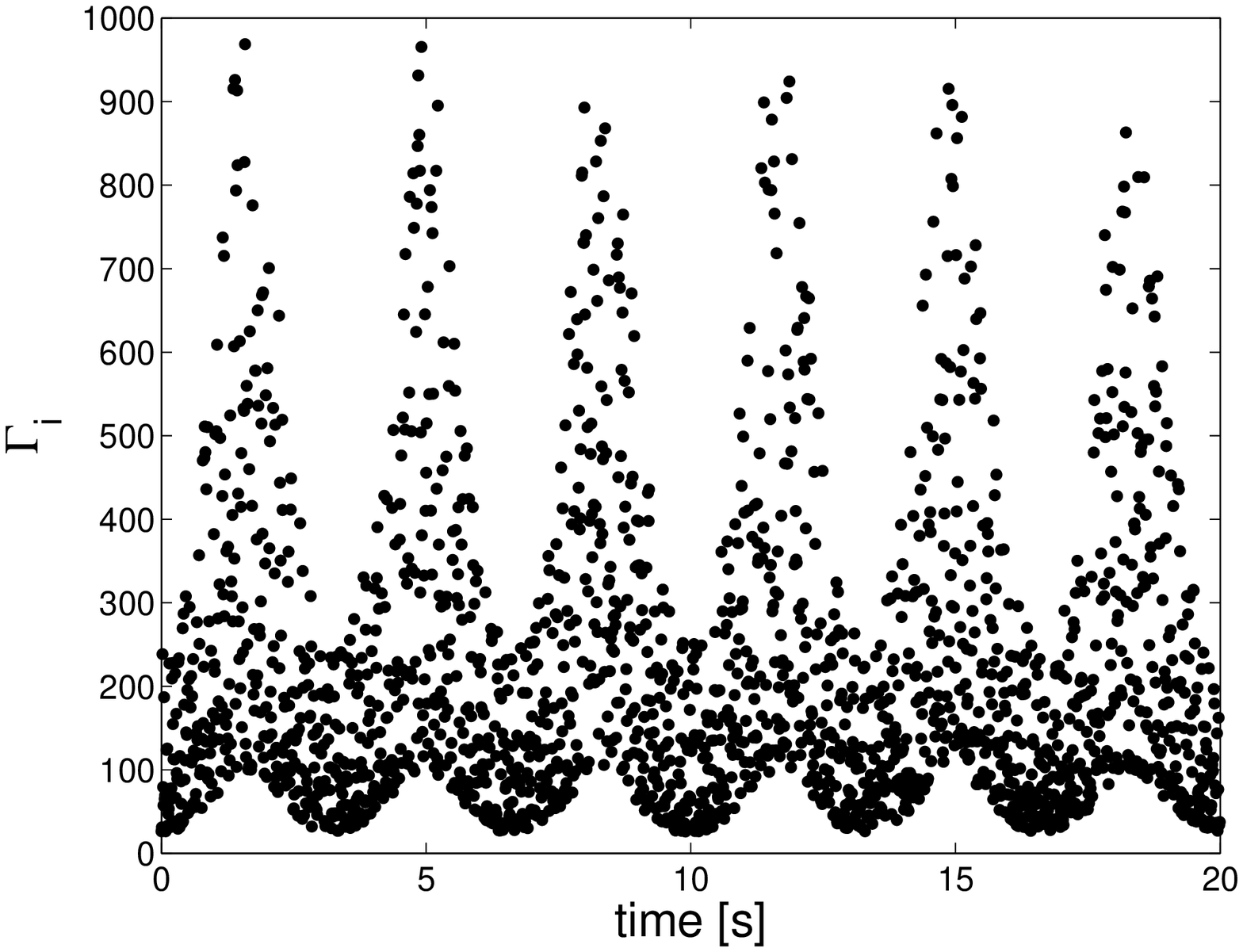}\\
\includegraphics[width=6.3cm]{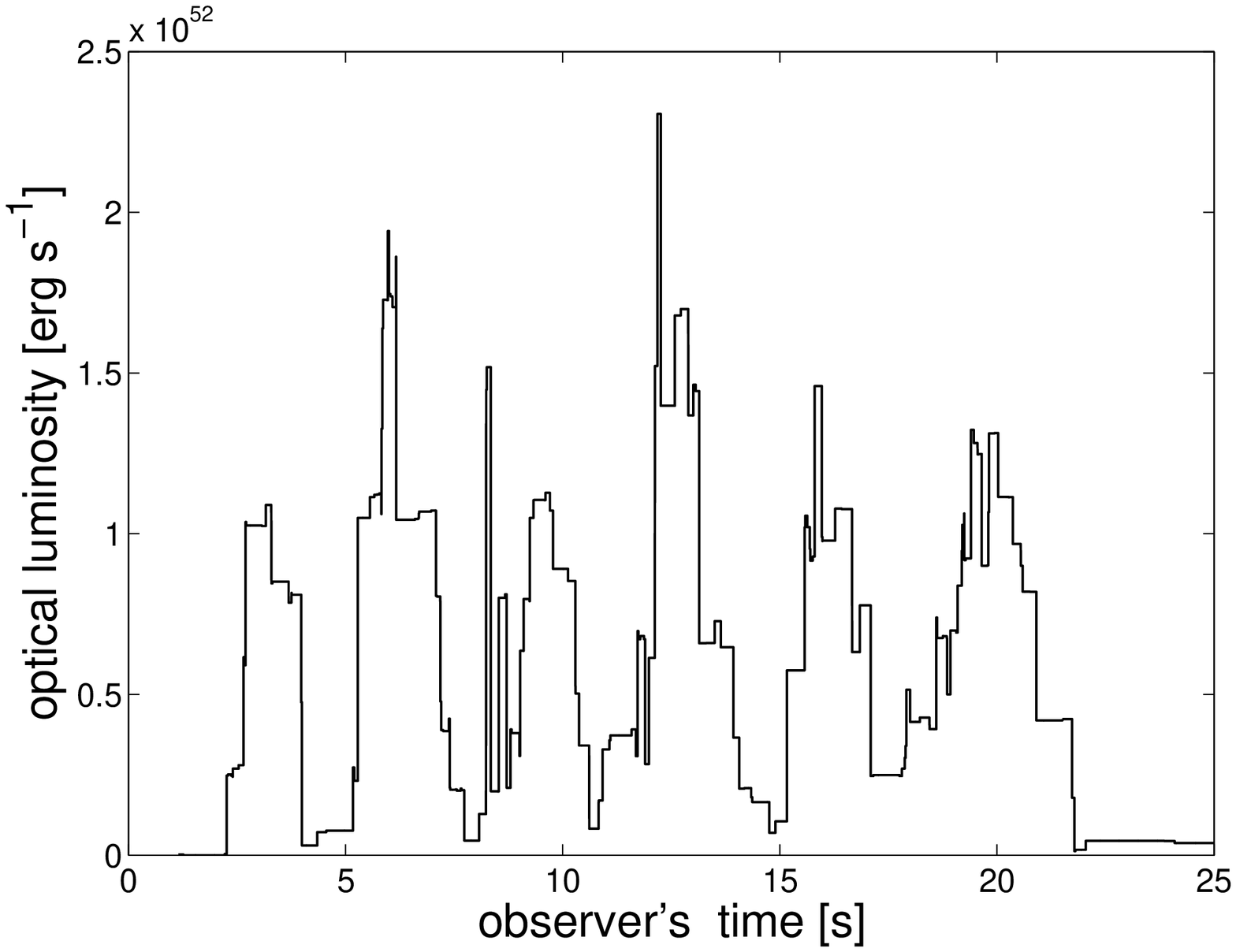}\\
\includegraphics[width=6.3cm]{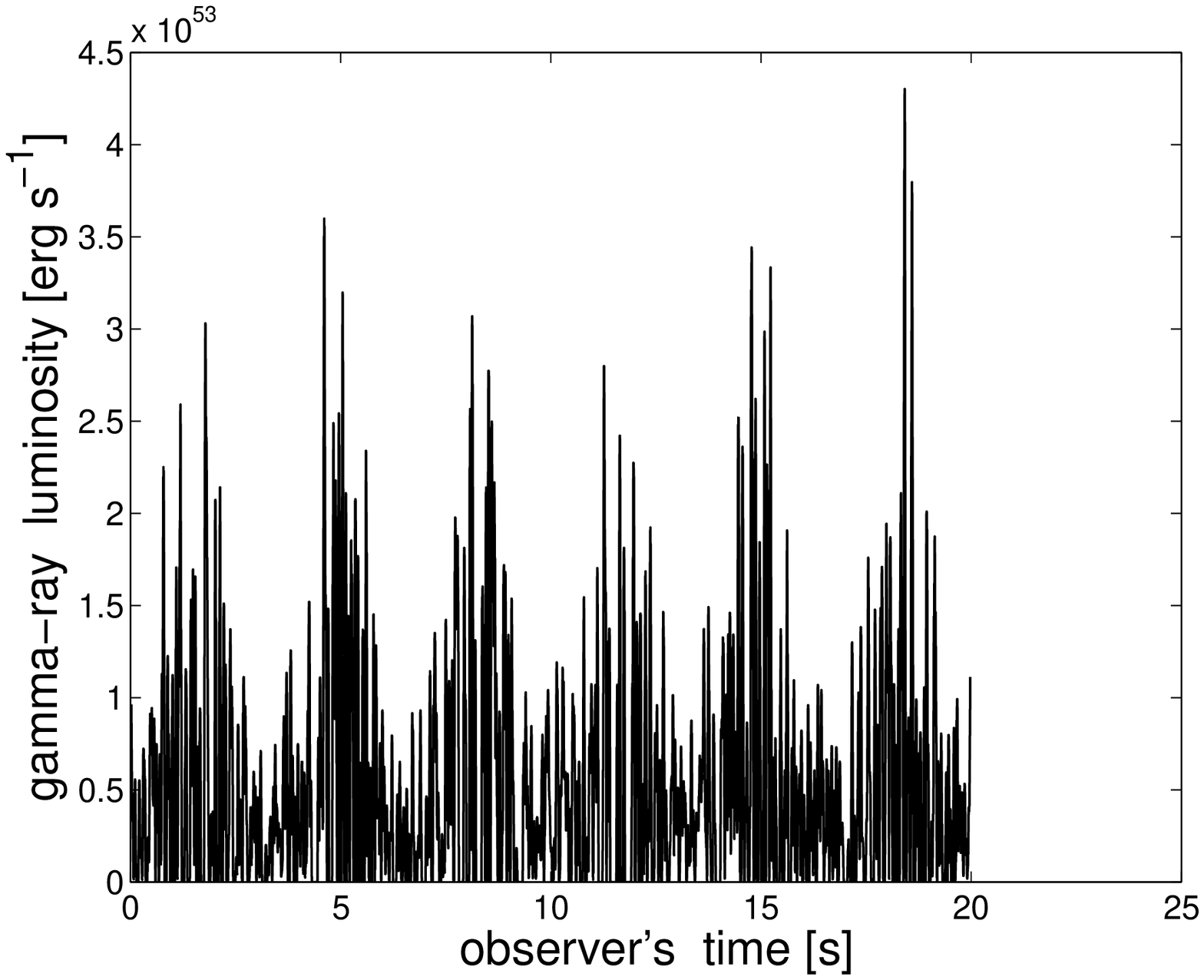}\\
\includegraphics[width=6.3cm]{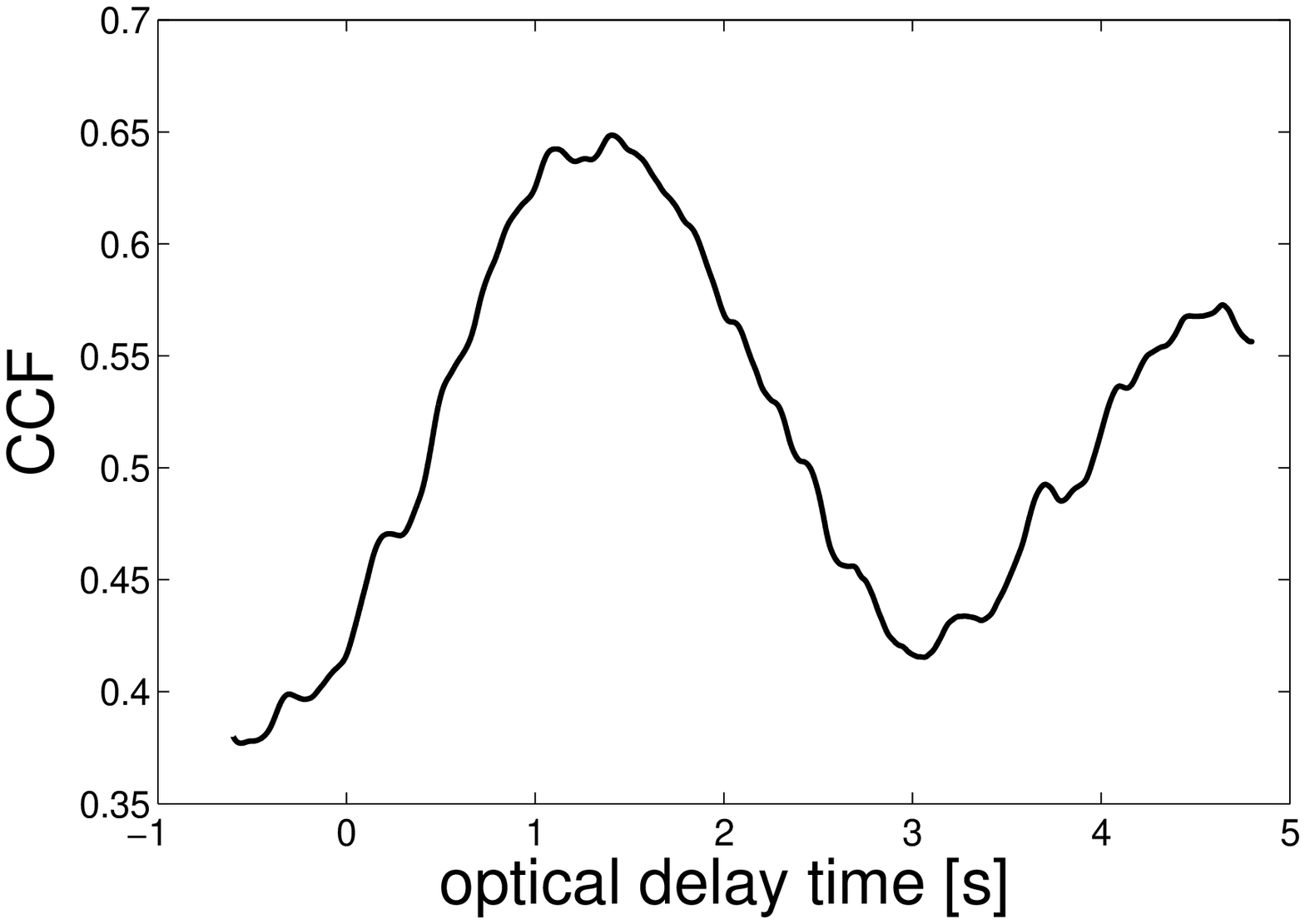}\\
\renewcommand{\captionlabelfont}{\bf}
   \captionsetup{labelsep=space}
   \caption{\small Results of Test 3. }\label{fig:test3}
\end{minipage}%
\begin{minipage}[t]{0.5\textwidth}
\centering
\includegraphics[width=6.5cm]{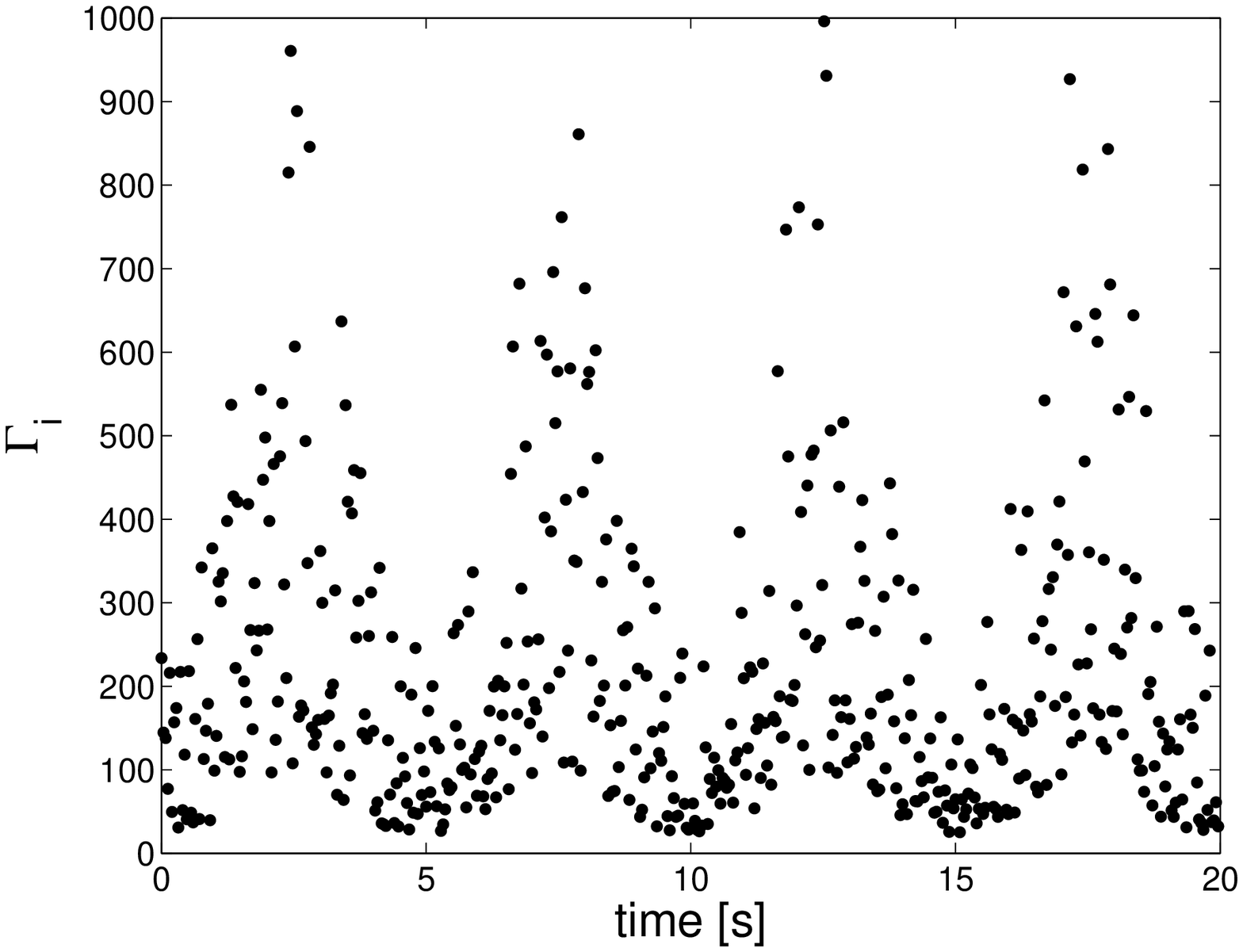}\\
\includegraphics[width=6.3cm]{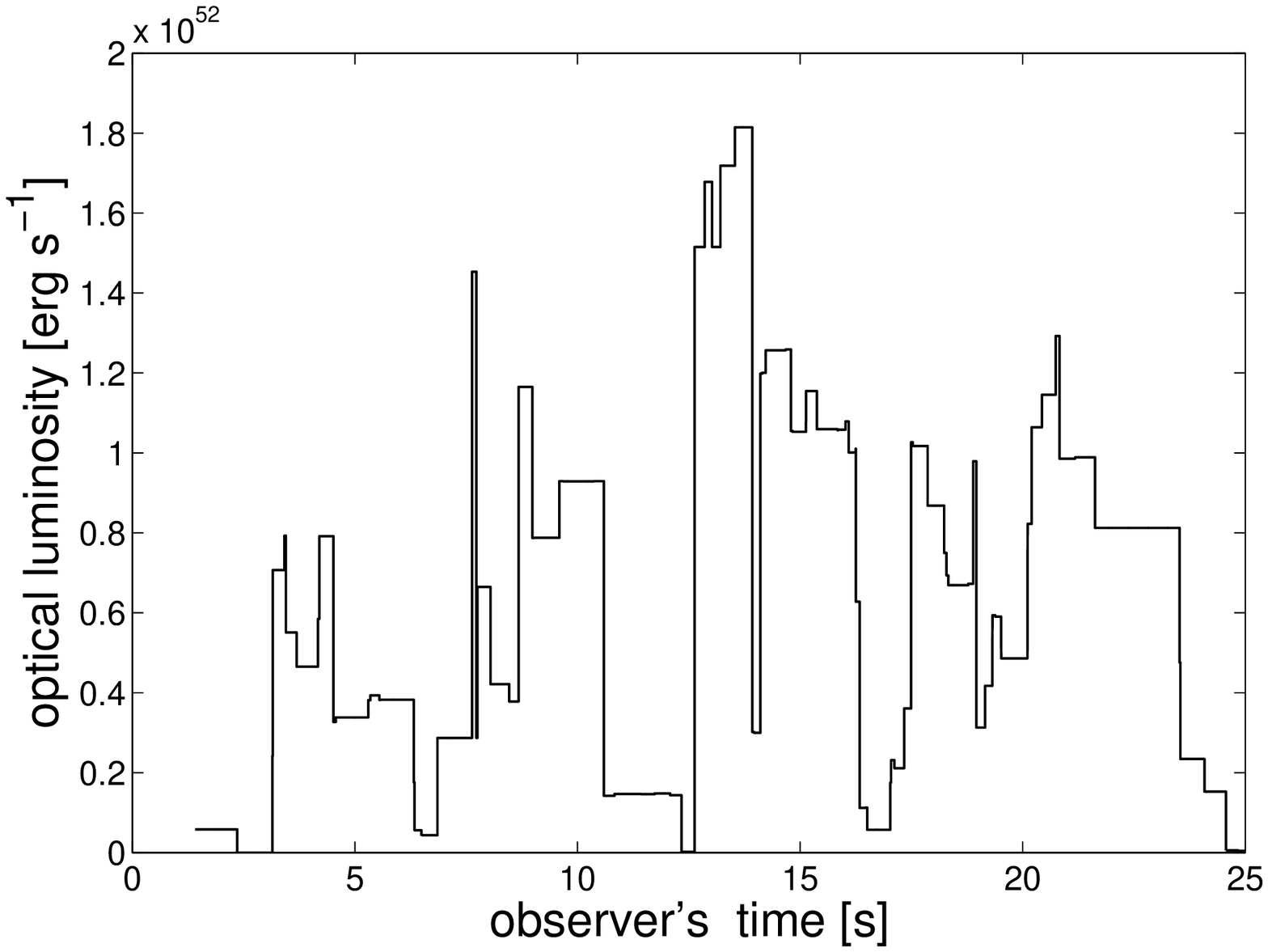}\\
\includegraphics[width=6.3cm]{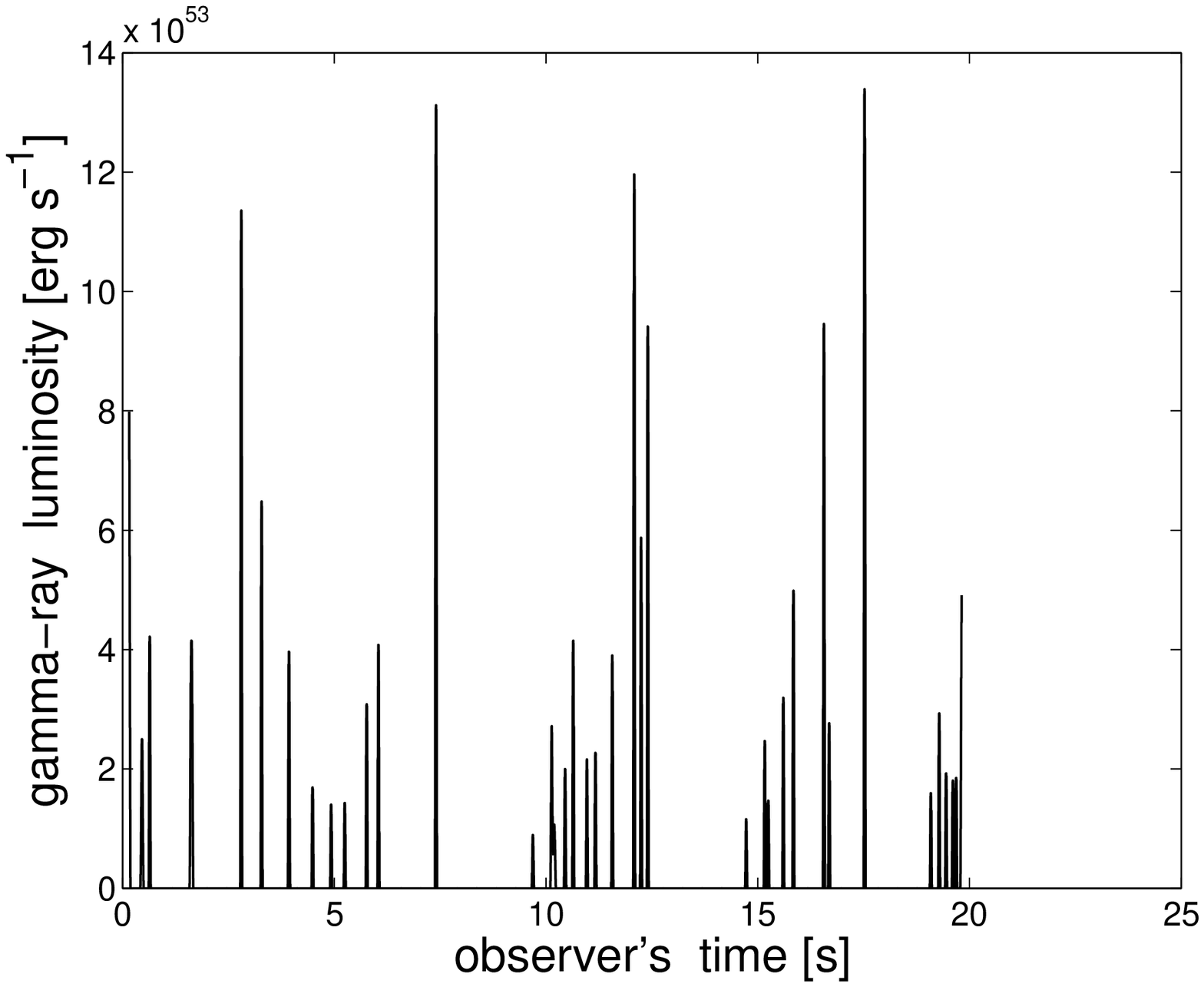}\\
\includegraphics[width=6.3cm]{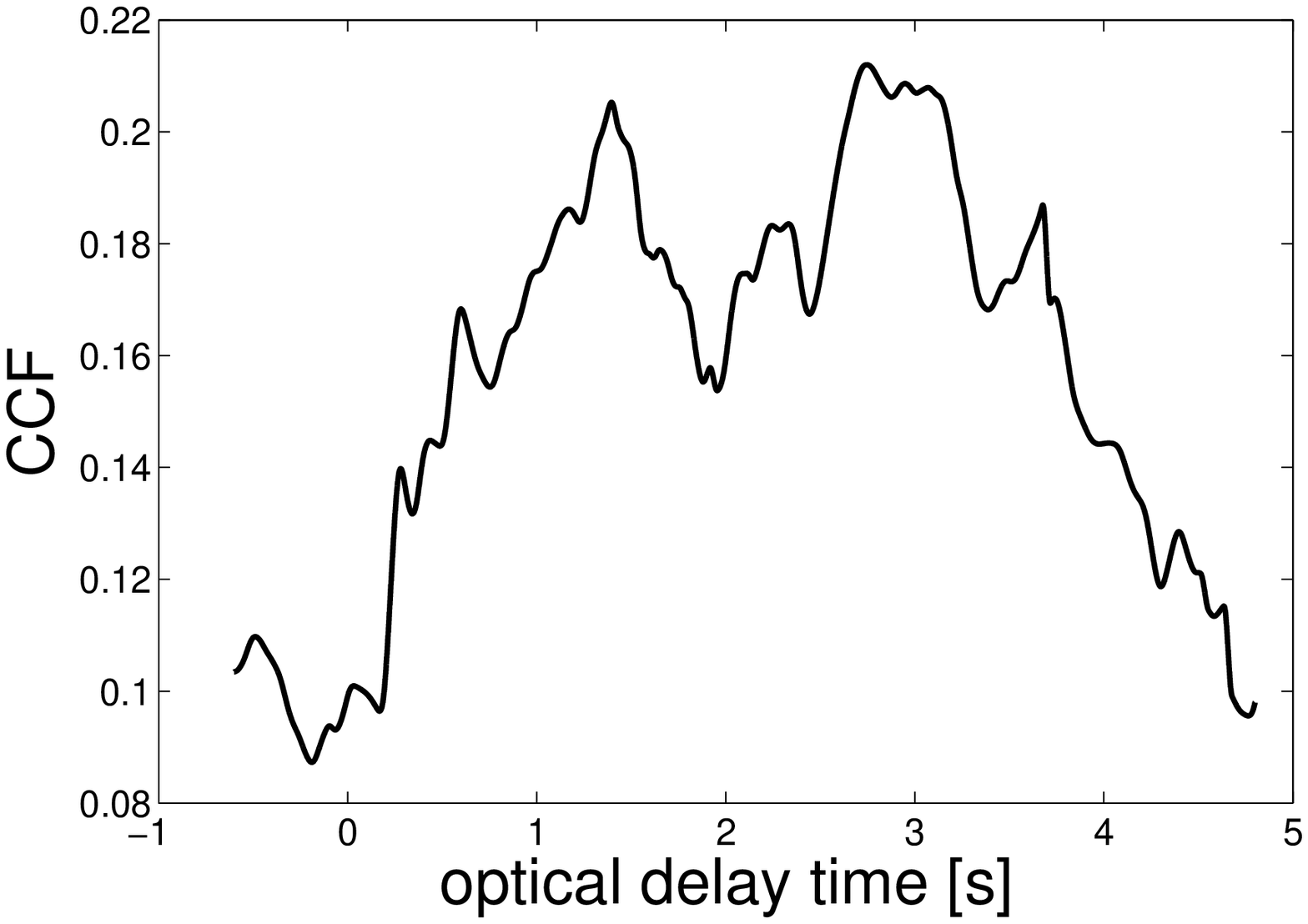}\\
\renewcommand{\captionlabelfont}{\bf}
   \captionsetup{labelsep=space}
   \caption{\small Results of Test 4.  }\label{fig:test4}
\end{minipage}
\end{figure}

\paragraph{Test 3.}
Next, we test the cases with different timescales of the  two
variability components of the initial LFs respectively. This is the
same as Test 2, except the period of the slow variability component
is changed to 3.3\,s. Fig.~\ref{fig:test3} shows the consequential
optical and gamma-ray lightcurves. Similar results to those
presented in Fig.~\ref{fig:test2} can be found for both lightcurves.
With narrower slow component duration ($\sim$ 3.3\,s), the
lightcurves contain the same number of periodic variability seen in
the initial LFs.

\paragraph{Test 4.}
We then increase the timescale of the rapid variability of  the
initial LFs by reducing the number of shells. Correspondingly, the
irregular short-scale variabilities in both lightcurves clearly have
larger timescales. We find the total energy emitted over the
gamma-ray band obviously decreases due to fewer collisions at small
radii (see Fig. \ref{fig:test4}). Fewer gamma-ray data
also lead to lower correlation between the two lightcurves.

\begin{figure}
\begin{minipage}[t]{0.5\textwidth}
\centering
\includegraphics[width=6.5cm]{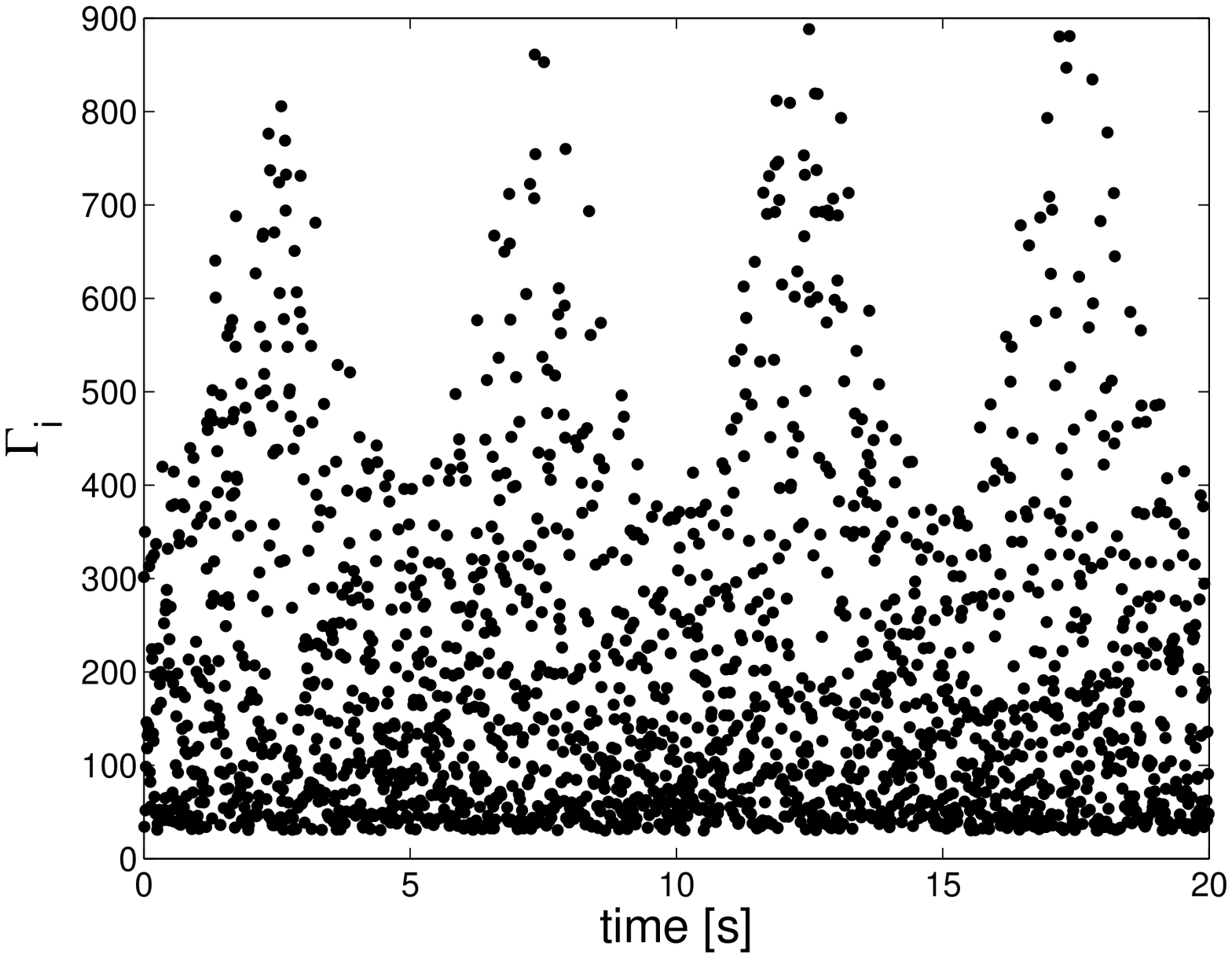}\\
\includegraphics[width=6.3cm]{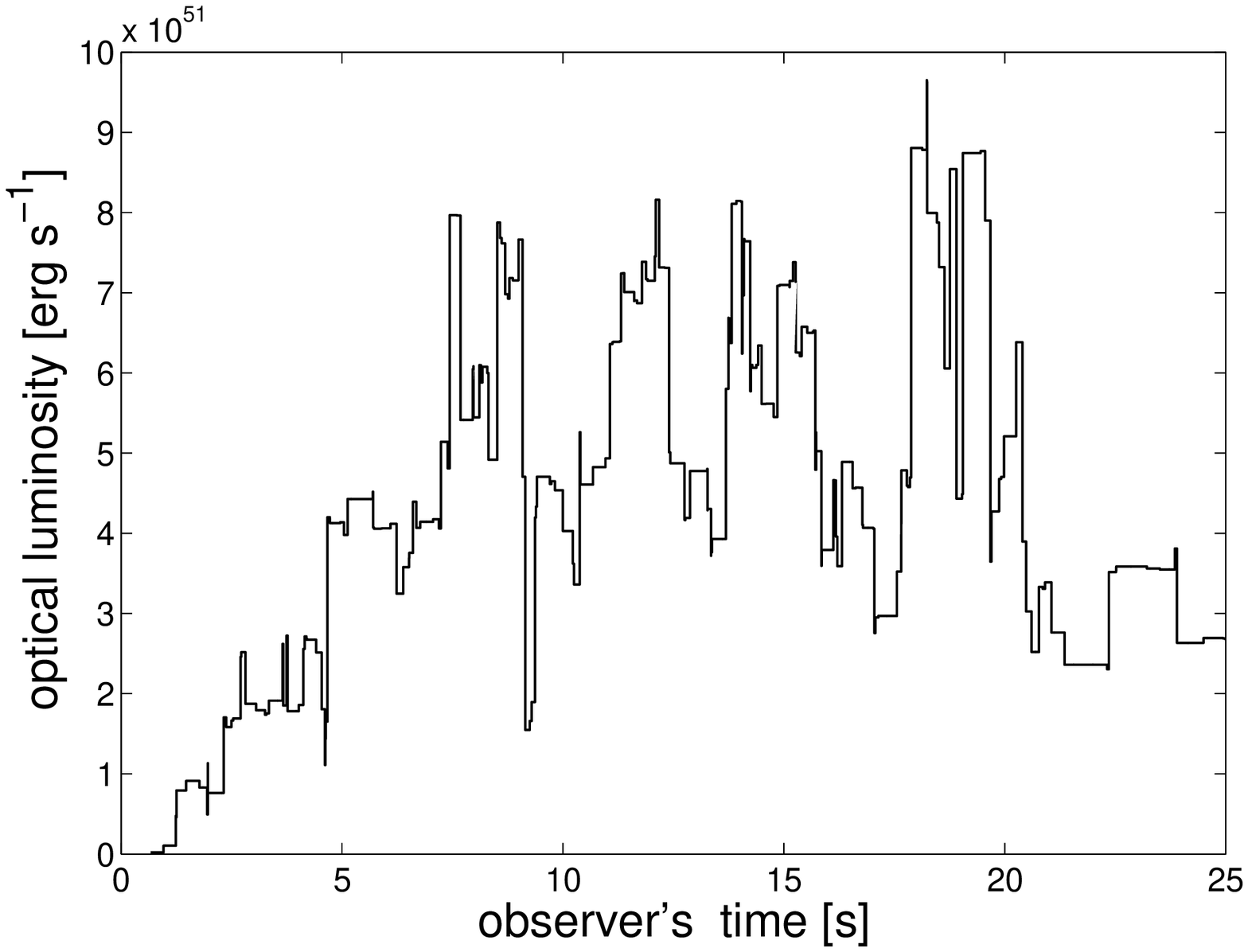}\\
\includegraphics[width=6.3cm]{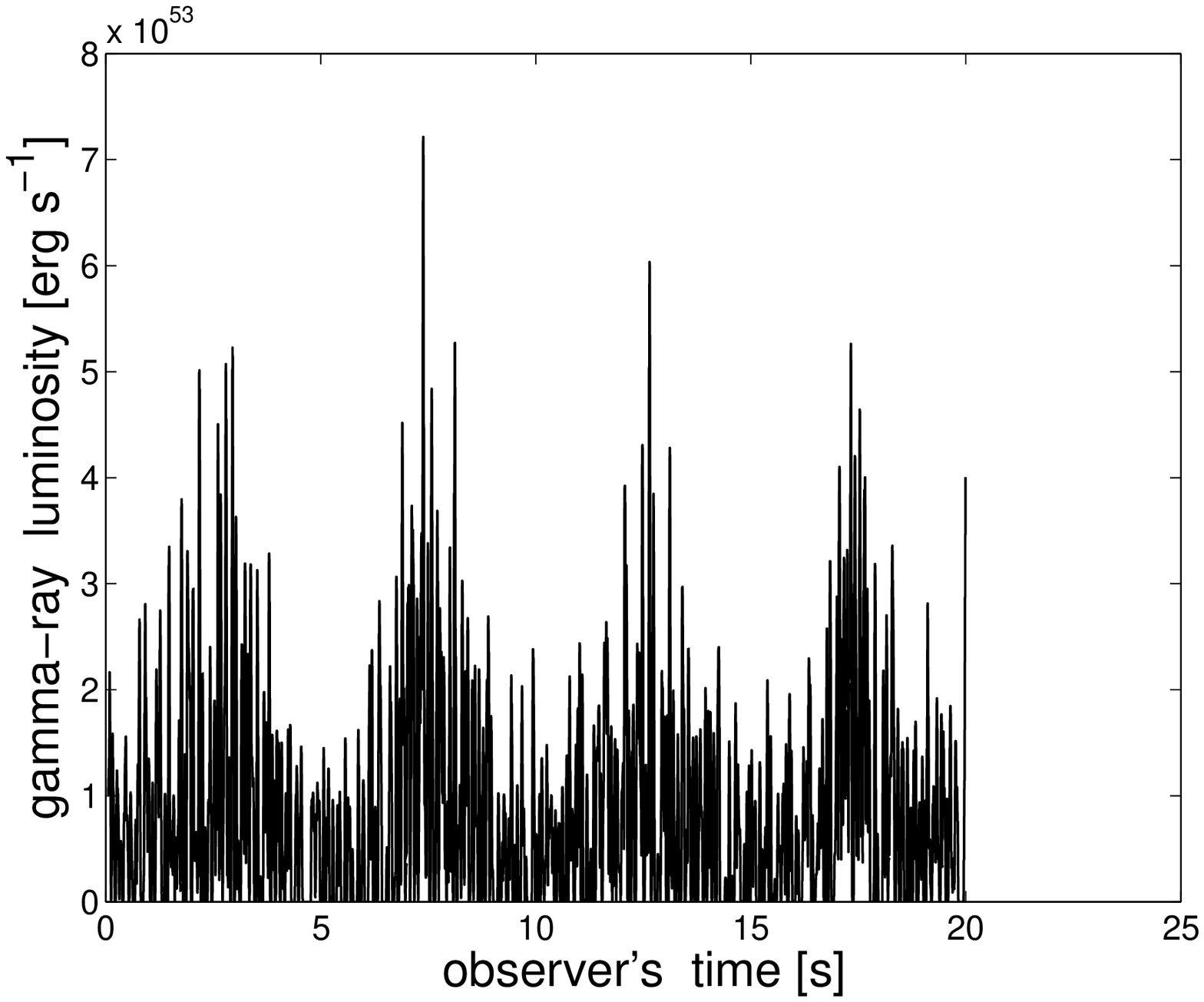}\\
\includegraphics[width=6.3cm]{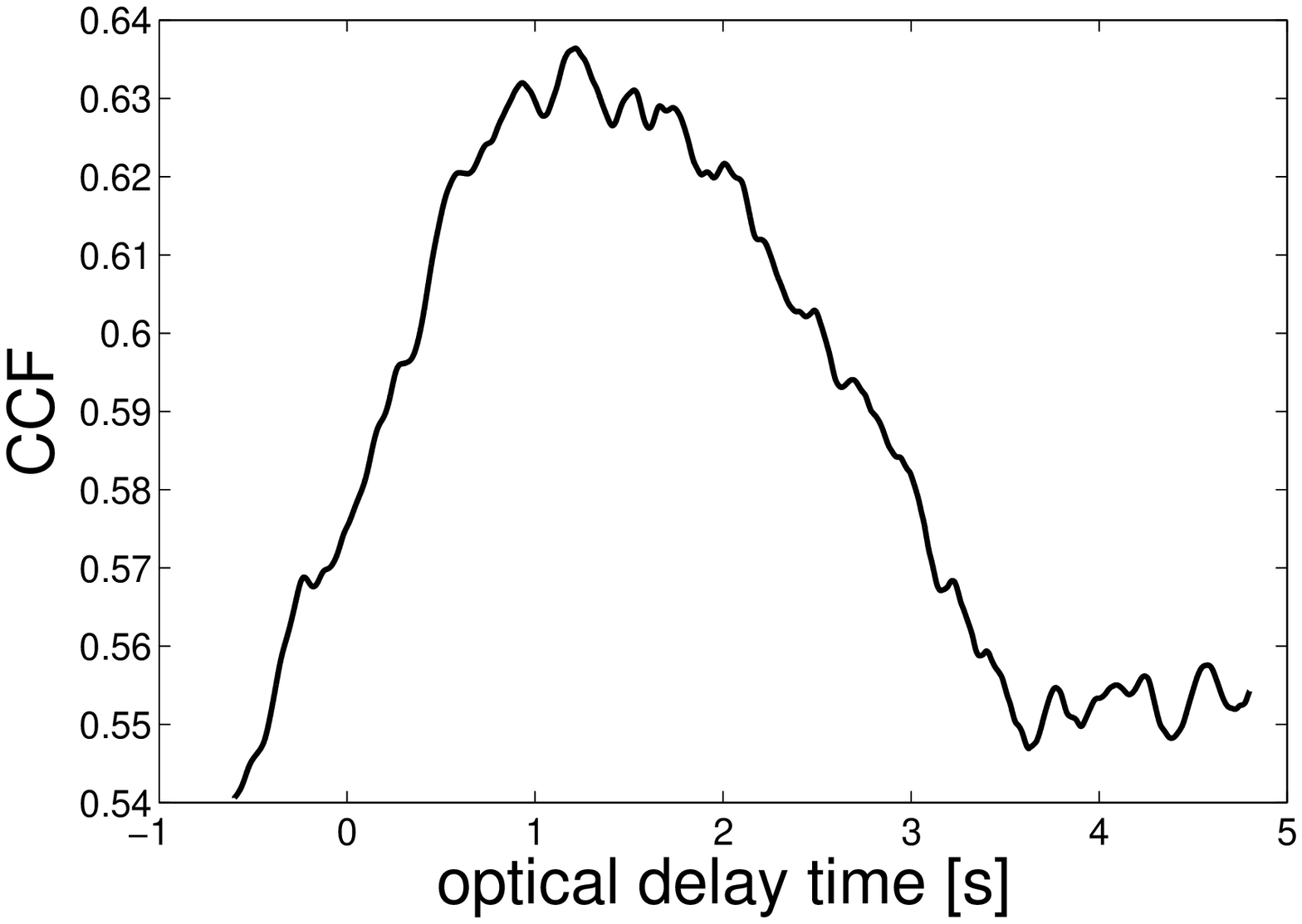}\\
\renewcommand{\captionlabelfont}{\bf}
   \captionsetup{labelsep=space}
   \caption{\small Results of Test 5, as a comparison with GRB 080319B (figure 2 in
   Beskin et al.~(\cite{Besk10})).}\label{fig:test5}
\end{minipage}
\end{figure}

Table 1 lists the parameters for the above simulation  tests.
Here we do not consider the redshift of the source, or all the
timescales in Table 1 should be increased by $1+\emph{z}$. As can be
seen from our simulation tests, the temporal behavior of prompt
optical and gamma-ray emission is sensitive to the changes of the
initial velocity variations inside the ejecta. The variability
features exhibited in lightcurves tend to strictly track the
temporal structure of the initial LFs of the shells.

\begin{table*}
\centering
\begin{threeparttable}
\caption[]{Parameters and results of the simulation tests.
}\label{Tab:par} \setlength\tabcolsep{2.3pt}
  \begin{tabular}{cccccccccc}
      \toprule
       & $N$\tnote{1} & $\tau_1$\,(s)\tnote{2}  & $\tau_2$\,(s)\tnote{3} & $r$\tnote{4} & $\Delta t$\,(s)\tnote{5} & $\delta t_o$\,(s)\tnote{6}&  $\delta t_{o,1}$\,(s)\tnote{7} & $\delta t_{o,2}$\,(s)\tnote{8} & $\delta t_\gamma$\,(s)\tnote{9}  \\
       \midrule
      Test 1 & 2000 &  0.01     & $-$ & 0.56 & 0.8 & 2.8  & 1.4    &   12.9    &    0.004  \\
      Test 2 & 2000 & 0.01    & 5 & 0.50 & 1.9 & 3.0 & 1.3     &   15.0    &    0.004   \\
      Test 3  & 2000 & 0.01 & 3.3 & 0.65 & 1.4 & 2.3  & 1.4    &   13.5    &    0.006   \\
      Test 4  & 500  & 0.04    & 5  & 0.21 &  2.7 & 6.0  & 1.9   &   22.0    &    0.012   \\
      Test 5  & 2000 &  0.01   & 5 & 0.64 & 1.2  & 2.8  & 1.7   &   11.7    &   0.005  \\
         \bottomrule
    \end{tabular}
       \begin{tablenotes}
       \footnotesize
        \item[1] The number of shells.
        \item[2] The timescales of the fast variability components existing in initial LFs.
        \item[3] The timescales of the slow variability components existing in initial LFs.
        \item[4] The highest correlation coefficient between optical and gamma-ray lightcurves.
        \item[5] The time lag between optical and gamma-ray emission.
       \item[6] The average timescale of variability in optical
       emission.
       \item[7] The average timescale of variability in optical emission for $t_{\rm obs} \leq 20$~s.
       \item[8] The average timescale of variability in optical emission for $t_{\rm obs} >20$~s.
       \item[9] The average timescale of variability in gamma-ray
       emission.
    \end{tablenotes}
   \end{threeparttable}
\end{table*}

\section{Comparison with GRB 080319B}
\label{sect:com} The high temporal resolution detection of  the
prompt optical emission of GRB 080319B and its gamma-ray counterpart
serve as the only available observed data to test our model.
Periodic variability on a few seconds time scale may exist in both
optical and rebinned gamma-ray lightcurves during the prompt phase
of the emission (Beskin et al.~\cite{Besk10}). Besides the four similar peaks,
short time-scale variability can be seen in the realistic
lightcurves, including the rapid optical variability on time scales
from several seconds to subseconds and a large amount of stochastic
variability in the gamma-ray emission (Beskin et al.~\cite{Besk10}).

In align with the observations, the  simulated lightcurves capture
both the underlying equidistant broad component and the short-scale
variability features. The detected time delay ($\sim$ 2\,s in the observer frame,
Beskin et al.~\cite{Besk10}) between the optical and gamma-ray emission has the
same order of magnitude as $\Delta$ \emph{t} shown in Table 1 (after
multiplied by $1+\emph{z}$). Despite the time lag, a clear
similarity exists between the optical and gamma-ray lightcurves, as
indicated in Beskin et al.~(\cite{Besk10}). This temporal correlation shows the
emission in the optical and gamma-ray ranges is generated by a
common mechanism, but at different locations.

Since our model can be applied to general GRB events by modifying
the initial Lorentz factor distribution of the ejected shells, in
order to construct the temporal structure of this specific burst, we
also present a slightly modified version of Test 2 (see Test 5 in
Table 1 and Fig. \ref{fig:test5}) as a comparison with figure 2 in
Beskin et al.~(\cite{Besk10}). Note that the initial LFs in Test 5
have the same variability timescales as those in Test 2, but the
adjusted Lorentz factor distribution leads to different lightcurves
characterized mainly by four overlapping peaks
(Fig.~\ref{fig:test5}) instead of separated pulses
(Fig.~\ref{fig:test2}).

\begin{figure}[htbp]
   \centering
   \includegraphics[width=0.7\textwidth]{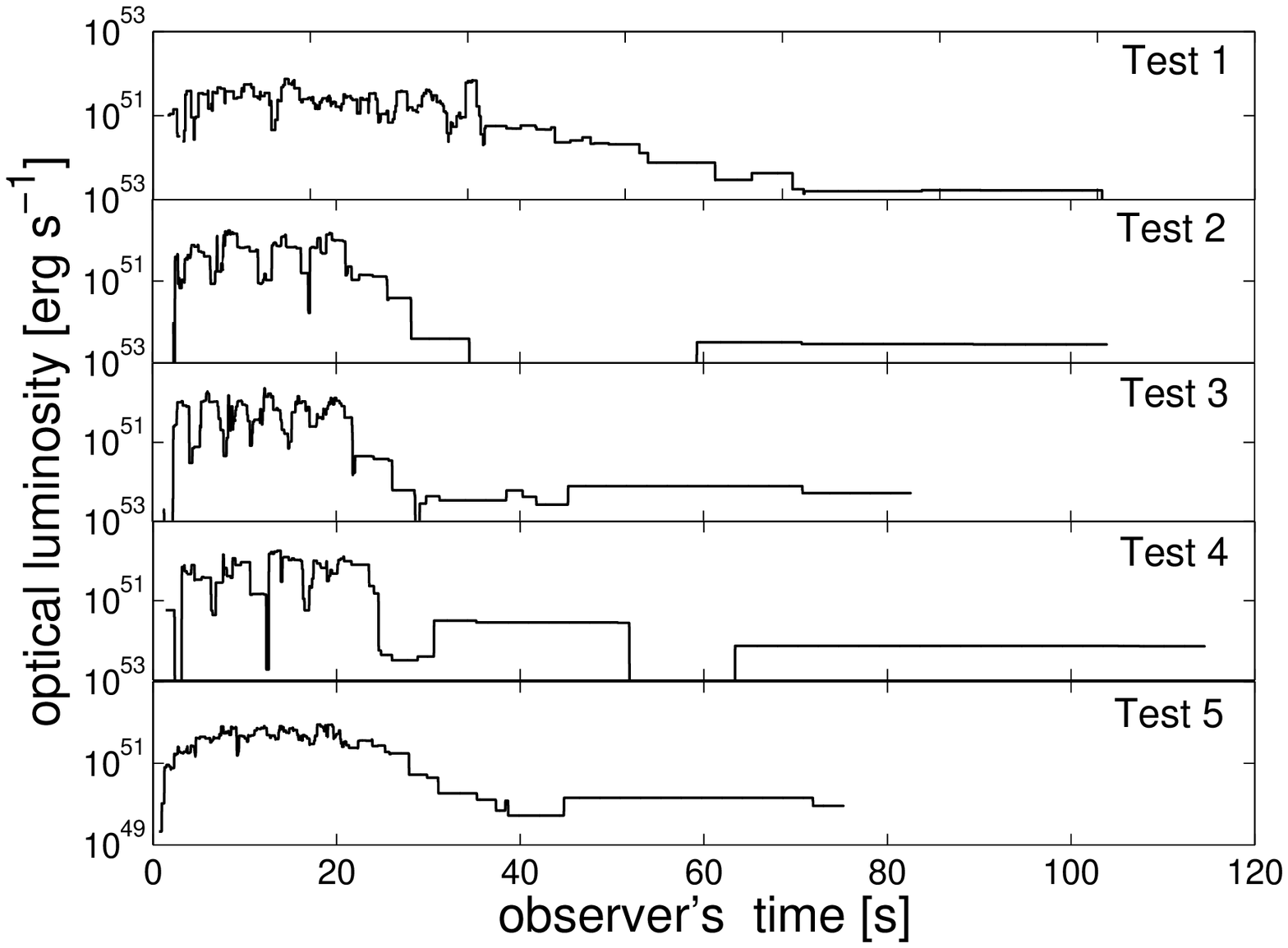}
   \renewcommand{\captionlabelfont}{\bf}
   \captionsetup{labelsep=space}
   \caption{\small The complete optical lightcurves from all simulation tests.}
   \label{Fig:5tgr}
   \end{figure}

\section{Discussion}
The simple model we proposed  is able to explain the basic temporal
structure of diverse GRB prompt emission. By adjusting the initial
velocity variations, i.e., the initial Lorentz factor distribution
in the outflow, our simulations can reproduce the complex
variability features in realistic lightcurves. One can predict the
highly variable temporal profile by controlling the initial variance
of the shell velocities, and more importantly, inspecting the
observational temporal features in the lightcurves of separate
energy bands allows us to assess the central engine activity in
detail.

The central engine activity and its detailed properties
regulate the temporal variations displayed in the sequence of shells
injected from the central engine, which are then reflected in the
observed variability components in the lightcurves. For GRB 080319B,
we require at least two timescales of the central engine activity to
reproduce the observed optical light curve, which is an important
implication of its central engine. In current situation we are not
able to identify what causes these two timescales. However,
following the common picture of the collapsar model a new black
hole, with a surrounding torus, is born in the center of the
progenitor star. For a black hole of $\sim3m_3M_\odot$, the
accretion time of material at the innermost radius $r_{\rm in}$ of
the torus is $t_{\rm acc}\sim2m_3$~ms (e.g., \cite{narayan01}),
comparable to the small timescale we require, whereas the accretion
time at the outer radius $r_{\rm out}\sim10^2r_{\rm in}$ is $t_{\rm
acc}\sim4m_3(r_{\rm out}/10^2r_{\rm in})^{3/2}$s (\cite{narayan01}),
similar to the larger timescale. This may imply that the material
supply at the outer edge of the torus is non-stable.

An important feature shown in our simulation results is the
time delay between the optical and gamma-ray lightcurves due to the
larger radius of the optical emission region. The delay time is
about 1~s in all the tests. The optical delay has been predicted by
Li \& Waxman (\cite{lw08}) in the single timescale case. They show
that for optical emission to avoid the synchrotron self absorption,
the outflow needs to expand to larger radii of
$R\sim10^{15}R_{15}$cm, and hence the optical delay time is $\Delta
t\sim
R/2\langle\Gamma\rangle^2c\sim1R_{15}(\langle\Gamma\rangle/10^2)^{-2}$s,
where $\langle\Gamma\rangle$ is the average value of initial Lorentz
factors of the ejected shells. This is comparable to the resulted
delay time in the tests. However, since many factors can influence
the delay time, more systematic simulations are required to
investigate the dependence of the delay time on the input parameters
in the future work.

\section{Conclusions}
\label{sect:conc} Starting with a two-component Lorentz  factor
distribution of the shells injected from the central engine, our
simulations generated the optical and gamma-ray lightcurves both as
a superposition of two variability components with different time
scales. The slow component has the same time scale as that exhibited
in the initial LFs, while the time scale of the fast component has a
trend with energy: the gamma-ray lightcurve has much more and faster
short-scale variabilities than its optical counterpart. Moreover,
the time scale of the fast variability changes with time in the
simulated optical lightcurve. The value in the first 20\,s of the
lightcurve is one order of magnitude smaller than that of the rest
part, which is in agreement with the finding in \cite{Marg08}.

The similarity between the optical and gamma-ray lightcurves, and
the time delay between them, provide strong evidence that the
emission has a common origin, but is generated at different radii
from the central engine. Further discussions show that the
lightcurves of prompt emission actually provide the temporal
information to clarify the physical nature of the central engine.
The different variability components seen in the lightcurves depend
on the intrinsic variability with different time scales of the
internal engine. Detailed temporal analysis of the simulated
lightcurves is necessary and will be performed in our future work.

Hopefully, more high temporal resolution observations of both prompt
optical emission from GRBs will be acquired by future telescopes,
e.g., UFFO-Pathfinder (\cite{uffo}) and SVOM-GWAC (\cite{gwac}).
These well-sampled bursts will further test the validity of our
model and provide a deeper insight into the source behavior.

\section*{Acknowledgement}
This work has been supported in part by the NSFC (11273005), the MOE
Ph.D. Programs Foundation, China (20120001110064), the CAS Open
Research Program of Key Laboratory for the Structure and Evolution
of Celestial Objects, and the National Basic Research Program (973
Program) of China (2014CB845800).

\section*{appendix}
We derive the width of the merged shell after shock crossings.
Consider two colliding shells, with the lab-frame width $\Delta_i$
and LF $\gamma_i$, where $i=1,2$. Shell 2 is faster, and overtakes
shell 1. There are double shocks and 4 regions in the interaction.
The unshocked shell 1 and 2 have number densities $n_1$ and $n_2$,
respectively, in their own rest frame. The densities in forward and
reverse shock regions are $n_f=4\gamma_fn_1$ and $n_r=4\gamma_rn_2$,
respectively, in their own rest frame, where $\gamma_{f,r}$ are the
LFs of the shocked fluid relative to unshocked shell 1 and 2,
respectively,
\begin{equation}
  \gamma_f=\gamma_1\gamma_{cm}(1-\beta_1\beta_{cm}),
  ~~\gamma_r=\gamma_2\gamma_{cm}(1-\beta_2\beta_{cm}).
\end{equation}
Note there is no relative motion between forward and reverse shock
regions.

Now consider everything in one frame, which is better to be the lab
frame. In the lab frame, the unshocked shell 1, shocked fluid (both
forward and reverse shock regions) and unshocked shell 2 have LFs
$\gamma_1$, $\gamma_{cm}$ and $\gamma_2$. So in the lab frame, the
densities of the 4 regions are, inner going, $\gamma_1n_1$,
$\gamma_{cm}n_f$, $\gamma_{cm}n_r$ and $\gamma_2n_2$. The
compression factor, i.e. the factor the density is enhanced by
shock, is
$k_f=\gamma_{cm}n_f/(\gamma_1n_1)=4\gamma_f\gamma_{cm}/\gamma_1$ for
forward shock region, or
$k_r=\gamma_{cm}n_r/(\gamma_1n_1)=4\gamma_r\gamma_{cm}/\gamma_2$ for
reverse shock region. Thus the shock-compressed, merged shell has a
width, right after merging without spreading, of
\begin{equation}
  \Delta_c=\frac{\Delta_1}{k_f}+\frac{\Delta_2}{k_r}=\frac{\Delta_1\gamma_1}{4\gamma_f\gamma_{cm}}+\frac{\Delta_2\gamma_2}{4\gamma_r\gamma_{cm}}
  =\frac1{4\gamma_{cm}^2}\left(\frac{\Delta_1}{1-\beta_1\beta_{cm}}+\frac{\Delta_2}{1-\beta_2\beta_{cm}}\right).
\end{equation}

\label{lastpage}
\end{document}